\documentclass[aps,prl,twocolumn,superscriptaddress,amsfont,graphicx,nofootinbib,preprintnumbers]{revtex4-1}%
\usepackage{color,graphicx,epsfig}
\usepackage{ifpdf}
\usepackage{amsmath}
\usepackage{bm}
\usepackage[english]{babel}
\usepackage{graphicx}%
\usepackage{amsfonts}%
\usepackage{amssymb}
\usepackage{braket}
\usepackage{hyperref}
\usepackage{enumerate}

\usepackage[T1]{fontenc}
\usepackage{slashed}
\usepackage[utf8]{inputenc}
\usepackage{xspace}
\usepackage{ulem}
\usepackage{array}
\usepackage{ulem,fancyvrb}
\usepackage{xcolor}
\newcommand{\tabincell}[2]{\begin{tabular}{@{}#1@{}}#2\end{tabular}}

\definecolor{nicered}{rgb}{0.7,0.1,0.1}
\definecolor{nicegreen}{rgb}{0.1,0.5,0.1}
\hypersetup{colorlinks,citecolor= nicegreen,linkcolor= nicered}

\usepackage{dcolumn}
\usepackage{bm}
\usepackage{slashed}

\begin{document}

\title{Photon-jet events as a probe of axion-like particles at the LHC}

\author{Daohan Wang}
\email{wangdaohan@itp.ac.cn}
\affiliation{Department of Physics and Institute of Theoretical Physics, Nanjing Normal University, Nanjing, 210023, China}
\affiliation{CAS Key Laboratory of Theoretical Physics, Institute of Theoretical Physics,
Chinese Academy of Sciences, Beijing 100190, China}
\affiliation{School of Physical Sciences, University of Chinese Academy of Sciences,
Beijing 100049, China}

\author{Lei Wu}
\email{corresponding author: leiwu@njnu.edu.cn}
\affiliation{Department of Physics and Institute of Theoretical Physics, Nanjing Normal University, Nanjing, 210023, China}

\author{Jin Min Yang}
\email{jmyang@itp.ac.cn}
\affiliation{CAS Key Laboratory of Theoretical Physics, Institute of Theoretical Physics,
Chinese Academy of Sciences, Beijing 100190, China}
\affiliation{School of Physical Sciences, University of Chinese Academy of Sciences,
Beijing 100049, China}

\author{Mengchao Zhang}
\email{corresponding author: mczhang@jnu.edu.cn}
\affiliation{Department of Physics and Siyuan Laboratory, Jinan University, Guangzhou 510632, P.R. China}

\date{\today}

\begin{abstract}
Axion-like particles (ALPs) are predicted by many extensions of the Standard Model (SM). When ALP mass lies in the range of MeV to GeV, the cosmology and astrophysics will be largely irrelevant. In this work, we investigate such light ALPs through the ALP-strahlung process $pp \to V a (\to \gamma\gamma)$ at the 14 TeV LHC with an integrated luminosity of 3000 fb$^{-1}$ (HL-LHC). With the photon-jet algorithm, we demonstrate that our approach can probe the mass range of ALPs, which is inaccessible to previous LHC experiments. The obtained result can surpass the existing limits on ALP-photon coupling in the ALP mass range from 0.3 GeV to 10 GeV.
\end{abstract}
\pacs{Valid PACS appear here}
\maketitle


\section{Introduction}
Searching for new particles is one of the crutial tasks in the LHC experiment. Light pseudo-scalars, such as axion-like particles (ALPs), are theoretically well-motivated. They generically appear in models with the spontaneous breaking of a global symmetry~\cite{Peccei:1977hh,Weinberg:1977ma,Wilczek:1977pj,Kim:1979if} or in the compactifications of string theory~\cite{Svrcek:2006yi,Arvanitaki:2009fg,Cicoli:2012sz}. Besides, ALPs may have connections with electroweak phase transition~\cite{Ballesteros:2016euj} and play a key role in solving the hierarchy problem~\cite{Graham:2015cka}.

In general, the ALP masses and couplings to SM particles are independent parameters. When ALP masses are below the MeV scale, they are already subject to many constraints from cosmological and astrophysical observations~\cite{Raffelt:1990yz,Marsh:2015xka}.
Besides, such light ALPs can serve as cold dark matter (DM)~\cite{Preskill:1982cy,Abbott:1982af,Dine:1982ah} and be explored by various astrophysical and terrestrial anomalies~\cite{Arias:2012az,Jaeckel:2014qea,Athron:2020maw,Gao:2020wer}.
While for ALPs in the mass range of MeV to GeV, the above cosmological and astrophysical bounds will vanish. But ALPs can have sizable contributions to low-energy observables in particle physics. Recently, many works have been paid to searching for ALPs in intensity frontiers~\cite{Izaguirre:2016dfi,Dolan:2017osp,Bauer:2019gfk,Banerjee:2020fue,Gu:2021lni}.

On the other hand, the ALPs can be directly produced at high energy colliders~\cite{Mimasu:2014nea,Knapen:2016moh,Barrie:2016ntq,Bauer:2017ris,Brivio:2017ije,Bauer:2018uxu,Ebadi:2019gij,Wang:2020ips,Ren:2021prq}. A considerable region of parameter space of ALPs has been constrained by the LEP and LHC data. For instance, ALPs can be searched for from the processes $e^+e^- \to \gamma a(\to \gamma\gamma)$ and $Z \to a\gamma$ at LEP~\cite{Jaeckel:2015jla}. The non-resonant production process of ALPs, such as $pp \to ZZ$, has been proposed at the LHC~\cite{Gavela:2019cmq}. The rare decays of Higgs boson $h \to Za(\to \gamma\gamma)$ and $h \to a(\to \gamma\gamma)a(\to \gamma\gamma)$ have been proposed to probe the ALP coupling to the photon in terms of the ALP mass on a future complete LHC Run-3 dataset in the future~\cite{Bauer:2017nlg}.

\begin{figure}[ht]
\begin{center}
\includegraphics[width=5cm]{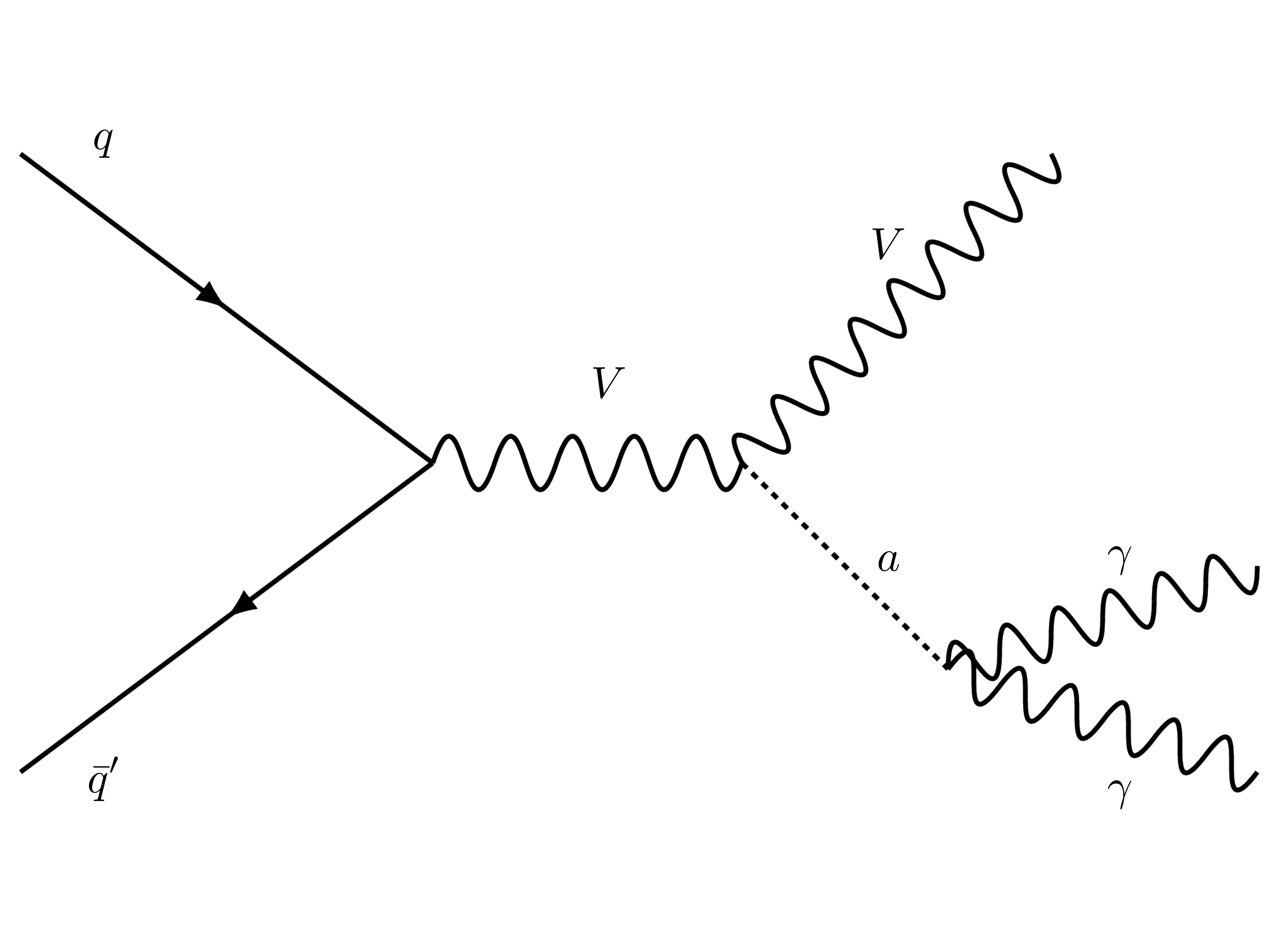}
\vspace{-0.5cm}
\caption{Feynman diagrams of ALP-strahlung process $pp \to Va(\to \gamma\gamma)$ at the LHC, where $V=W,Z$.}
\label{Feynman_diagrams}
\end{center}
\end{figure}
However, current LHC analyses are sensitive to the ALPs in the parameter region $m_a\gtrsim 10$ GeV. When ALPs become light, they can be highly boosted, and thus the two photons from the ALP decay are recognized as one object in the detector.
This will lead to an interesting signature ``photon-jet'' at the LHC.
In this work, we will use jet substructure variables to analyze these ``photon-jet'' events from the decay of ALPs, as the method proposed in Ref.~\cite{Ellis:2012sd,Ellis:2012zp}.
We will focus on ALPs that only couple to the electroweak vector bosons with a mass in MeV to GeV range. Such light electroweak ALPs have obtained an increasing amount of interest due to current collider data.
In this work, there is no ALP-gluon coupling so that the dominant production channel of electroweak ALPs at the LHC is ALP-strahlung process $pp \to Va$, where $V=W,Z$ (see Fig.~\ref{Feynman_diagrams}). We find that our proposal can extend the LHC sensitivity to parameter space unreachable in previous studies.

\section{Model and Photon-jet}
The relevant effective interactions of ALP with electroweak gauge bosons up to dimension-5 is given by~\cite{Brivio:2017ije}
\begin{eqnarray}
	\mathcal{L}_{\text{eff}}
	&\supseteq &
 \frac{1}{2} (\partial^\mu a)(\partial_\mu a) - \frac{1}{2} m_a^2 a^2-  C_{BB} \frac{a}{f_a} B_{\mu\nu} \tilde{B}^{\mu\nu}\nonumber\\
	& &
	-  C_{WW} \frac{a}{f_a} W^{i}_{\mu\nu} \tilde{W}^{\mu\nu,i},
\label{general-NLOLag-lin}
\end{eqnarray}
where the ALP field is denoted by $a$, and the field strengths for the SM gauge groups are  denoted as $V_{\mu\nu} \equiv \partial_\mu V_\nu - \partial_\nu V_\mu$ and $\tilde{V}_{\mu\nu} \equiv \epsilon_{\mu\nu\rho\sigma}V^{\rho\sigma}$. The $W_{\mu\nu}$ and $B_{\mu\nu}$ are field strengths for $SU(2)_L$ and $U(1)_Y$, respectively.
Both $C_{WW}$ and $C_{BB}$ contribute to the interaction of the ALP with two photons. The dimentionful coupling $g_{a\gamma\gamma}$,  $g_{aWW}$, $g_{aZZ}$, $g_{a\gamma Z}$ are given by
\begin{eqnarray}
	g_{a\gamma\gamma} &=& \frac{4}{f_a} (C_{BB} \cos\theta_W^2 + C_{WW} \sin\theta_W^2)\\
	g_{aWW} &=& \frac{4}{f_a} C_{WW}\\
	g_{aZZ} &=& \frac{4}{f_a} (C_{BB} \sin\theta_W^2 + C_{WW} \cos\theta_W^2)\\
	g_{a\gamma Z} &=& \frac{8}{f_a}\sin\theta_W\cos\theta_W (C_{WW} - C_{BB})
\end{eqnarray}
where $\theta_W$ is the Weinberg angle. For simplicity, we set $C_{WW} = C_{BB}$ in our study. We consider two ALP-strahlung processes $p p \to W^\pm(\to \ell^\pm \nu)a(\to \gamma\gamma)$ and $p p \to Z(\to \ell^+\ell^-)a(\to \gamma\gamma)$ as our signals. The dominant SM background processes include $V\gamma$, $Vj$, and QCD di-jets. In the LHC experiment, when $m_a \gtrsim 10$ GeV, the two photons from ALP decay can be separated enough and identified as $2\gamma$ events. On the other hand, if ALP mass is lighter than a few hundred MeV, those photons are highly collimated so that they will be detected as single photon events. While between these two mass limits, the two photons will be seen like a ``photon-jet''. In this case, we can use jet substructure techniques to discriminate photon-jets from single photons and QCD jets~\cite{Dobrescu:2000jt,Chang:2006bw,Toro:2012sv,Draper:2012xt,Ellis:2012sd,Ellis:2012zp}.

We implement the lagrangian of  Eq.~\ref{general-NLOLag-lin} in \textsf{FeynRules}~\cite{Alloul:2013bka} to generate the corresponding UFO model file. We use \textsf{MadGraph5 aMC@NLO}~\cite{Alwall:2014hca} to calculate the production cross sections and generate the signal and background events. Then the parton level events are showered and hadronized by \textsf{Pythia8}~\cite{Sjostrand:2014zea}. The detector simulation is performed by \textsf{Delphes}~\cite{deFavereau:2013fsa}. \textsf{FastJet}~\cite{Cacciari:2011ma} is used for jet clustering. The electron and muon identification efficiencies are taken as default values.  Based on the energy-flow algorithm~\cite{CMS:2009nxa}, we cluster EflowPhotons, EflowNeutralHadrons and ChargedHadrons into jets by using the anti-$k_t$ algorithm~\cite{Cacciari:2008gp} with $R_j=0.4$. Only the leading jet with $p_{T}$ > 50 GeV for each event is retained for further analysis. Then we recluster the rest energy deposits in each jet using $k_t$ algorithm~\cite{Catani:1993hr, Ellis:1993tq}, which determines a recombination tree for the jets. Besides, we also consider the pileup interaction effect. The low-$Q^2$ soft QCD pile-up events are generated by \textsf{Pythia8} and then simulated by \textsf{Delphes}. We take the default parametrization in the CMS card to distribute the minimum-bias pile-up events and hard scattering events in time and $z$ positions. The average amount of pileup events per bunch-crossing is considered to be 40. Besides, it should be noted that the calibration of jet energy scale is one of the main uncertainties in the jet analysis at the LHC. Since the traditional photon isolation criterion is not used, the new method of calibrating the photon-jets energy scale based on a full simulation of detector and real data is needed. This is challenging for a phenomenological study and is beyond the scope of our work.


To discriminate the signal events from the SM backgrounds, we use jet substructure algorithm to select our photon-jet events. We generate $p p \to Z(\to \nu\bar{\nu})a$, $p p \to Z(\to \nu\bar{\nu})j$ and $p p \to Z(\to \nu\bar{\nu})\gamma$ as the training samples of photon-jet, QCD jet, and single photon events, respectively. According to Ref.~\cite{Ellis:2012sd,Ellis:2012zp}, the following variables are used in our substructure analysis:
\begin{figure*}[ht]
\begin{center}
\includegraphics[width=8.4cm]{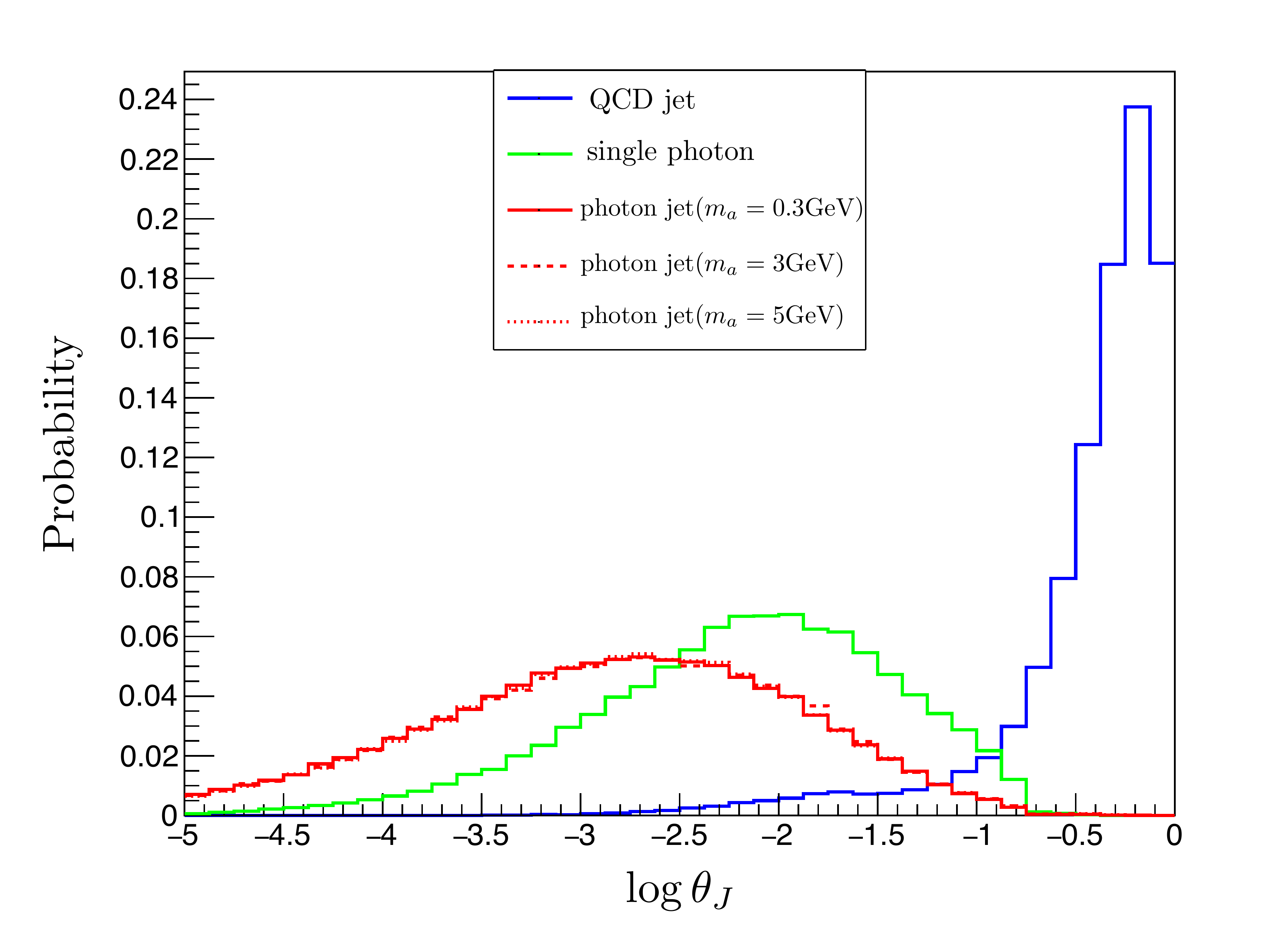}
\includegraphics[width=8.4cm]{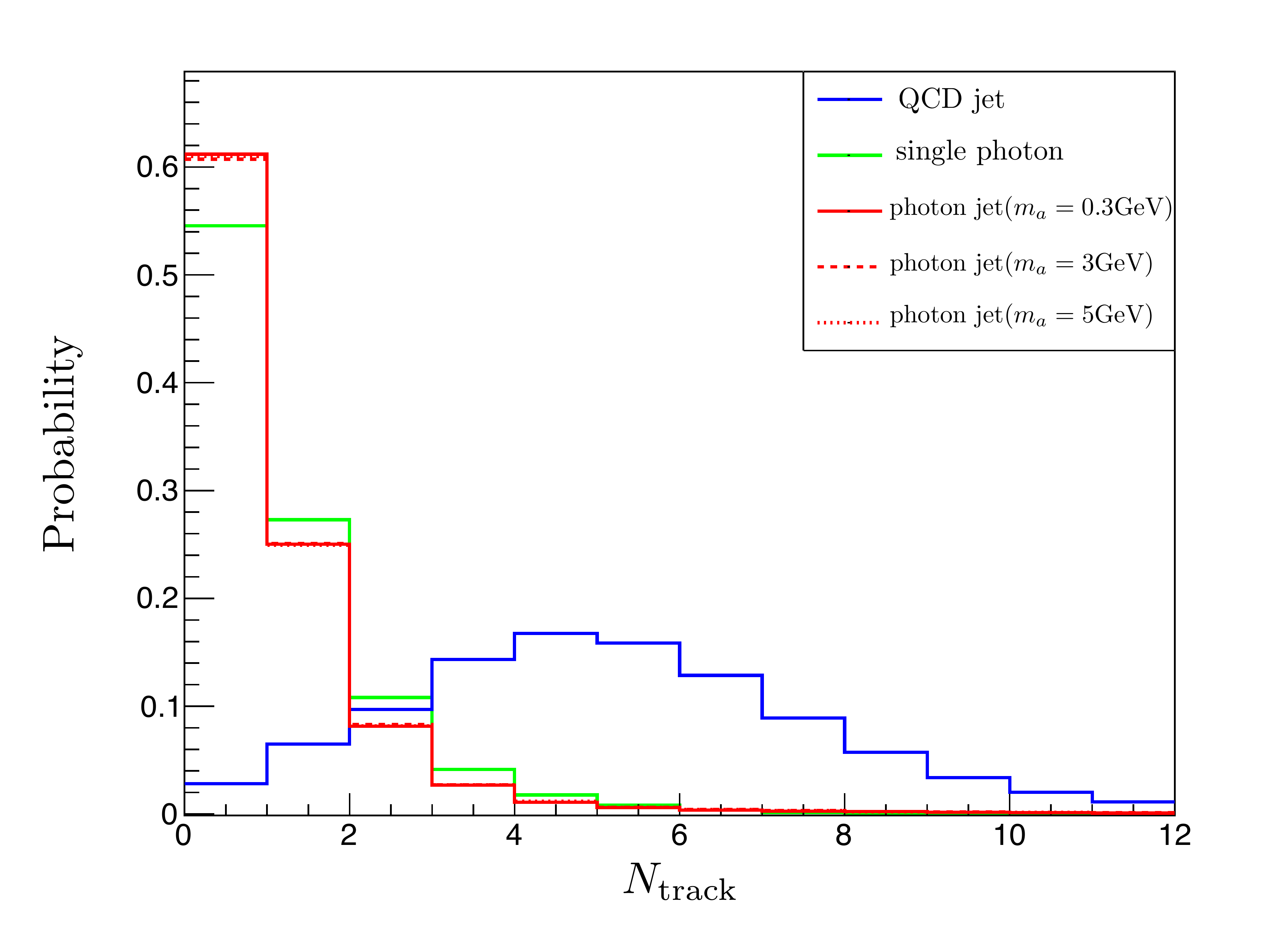}
\caption{Distributions of jet substructure variables (log$\theta_J$, $N_{\text{track}}$) for single photon, photon-jet and QCD jet events.}
\label{distributions1}
\end{center}
\end{figure*}
\begin{figure*}[ht]
\begin{center}
\includegraphics[width=8.4cm]{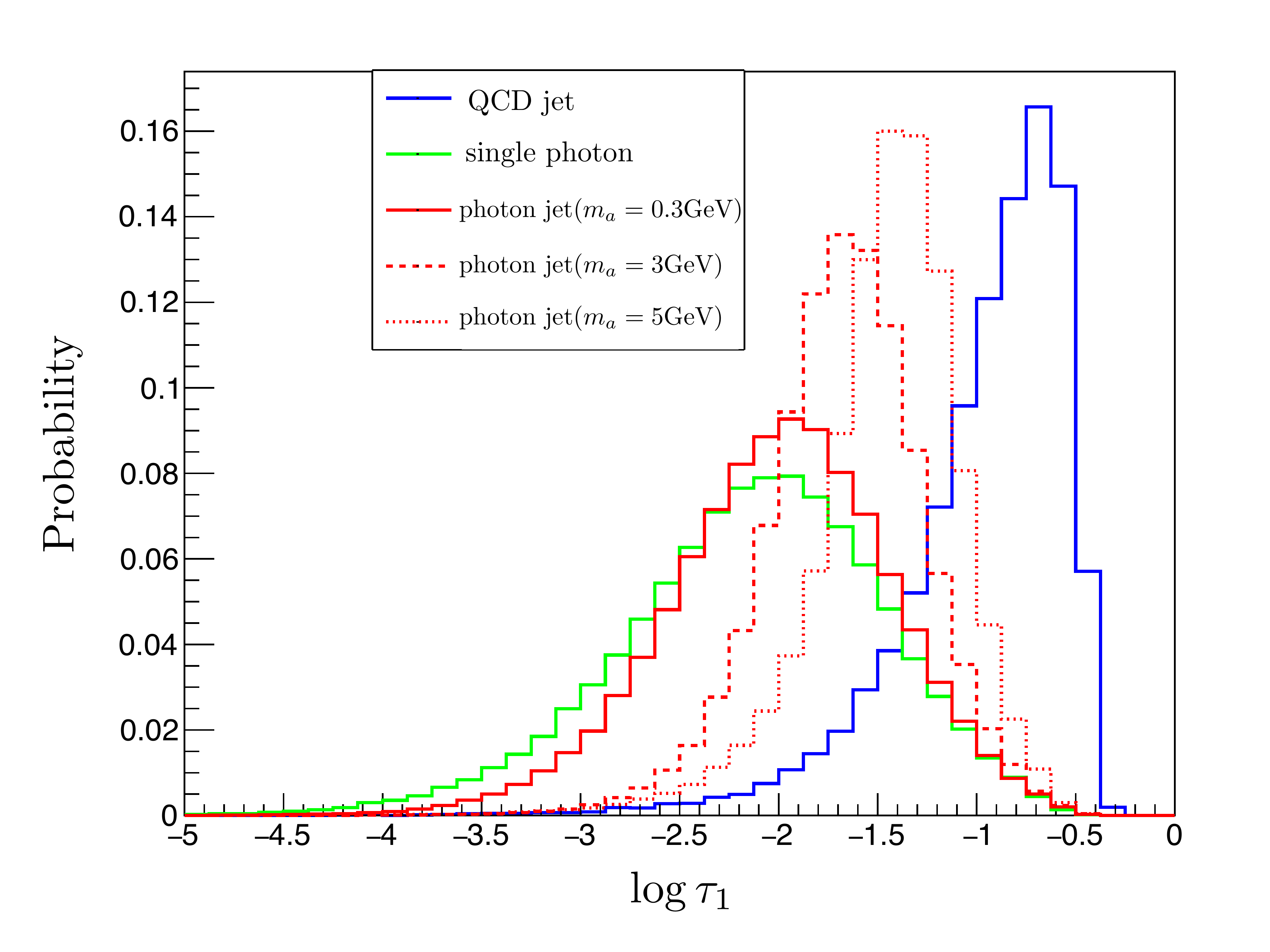}
\includegraphics[width=8.4cm]{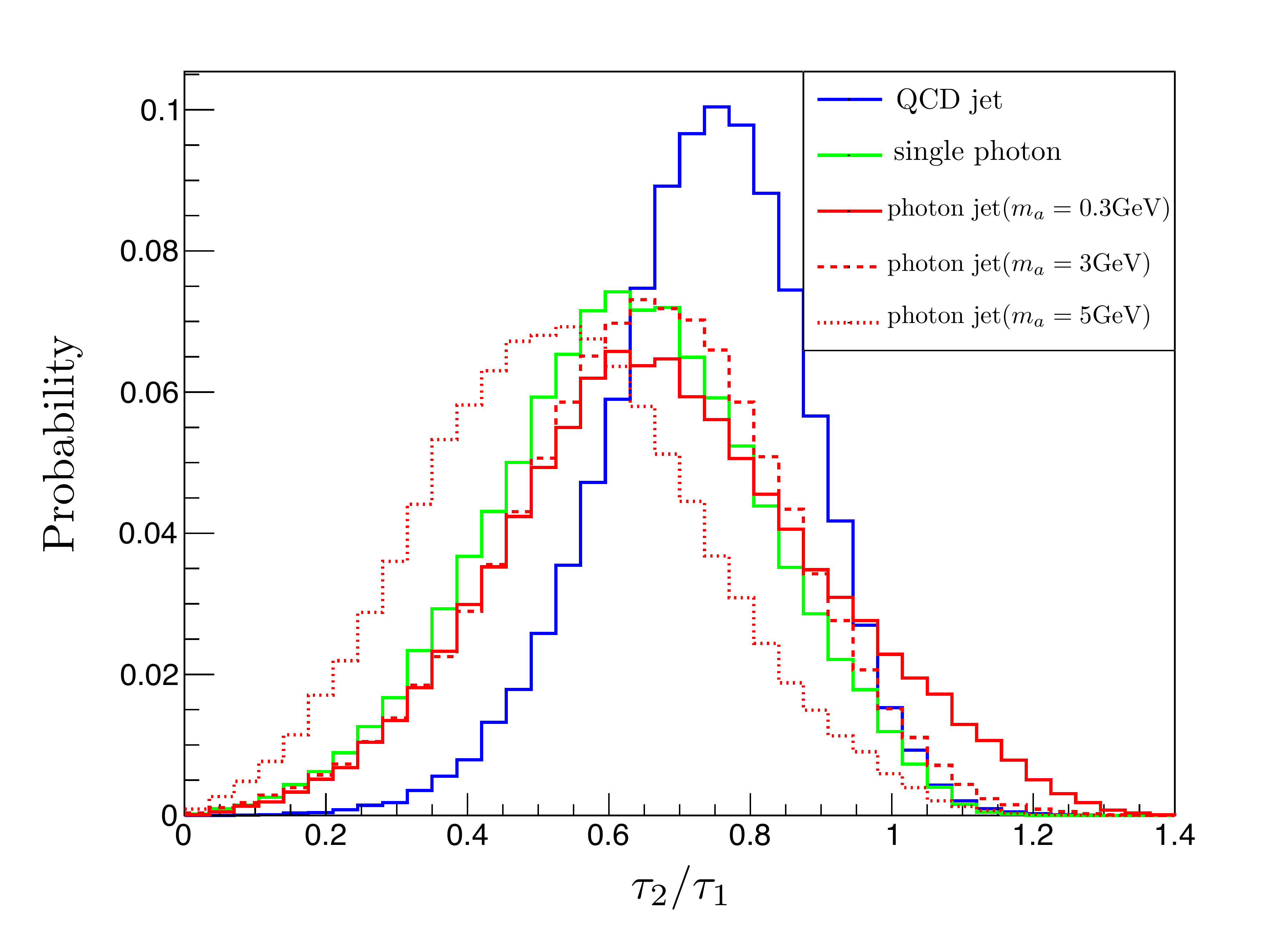}
\includegraphics[width=8.4cm]{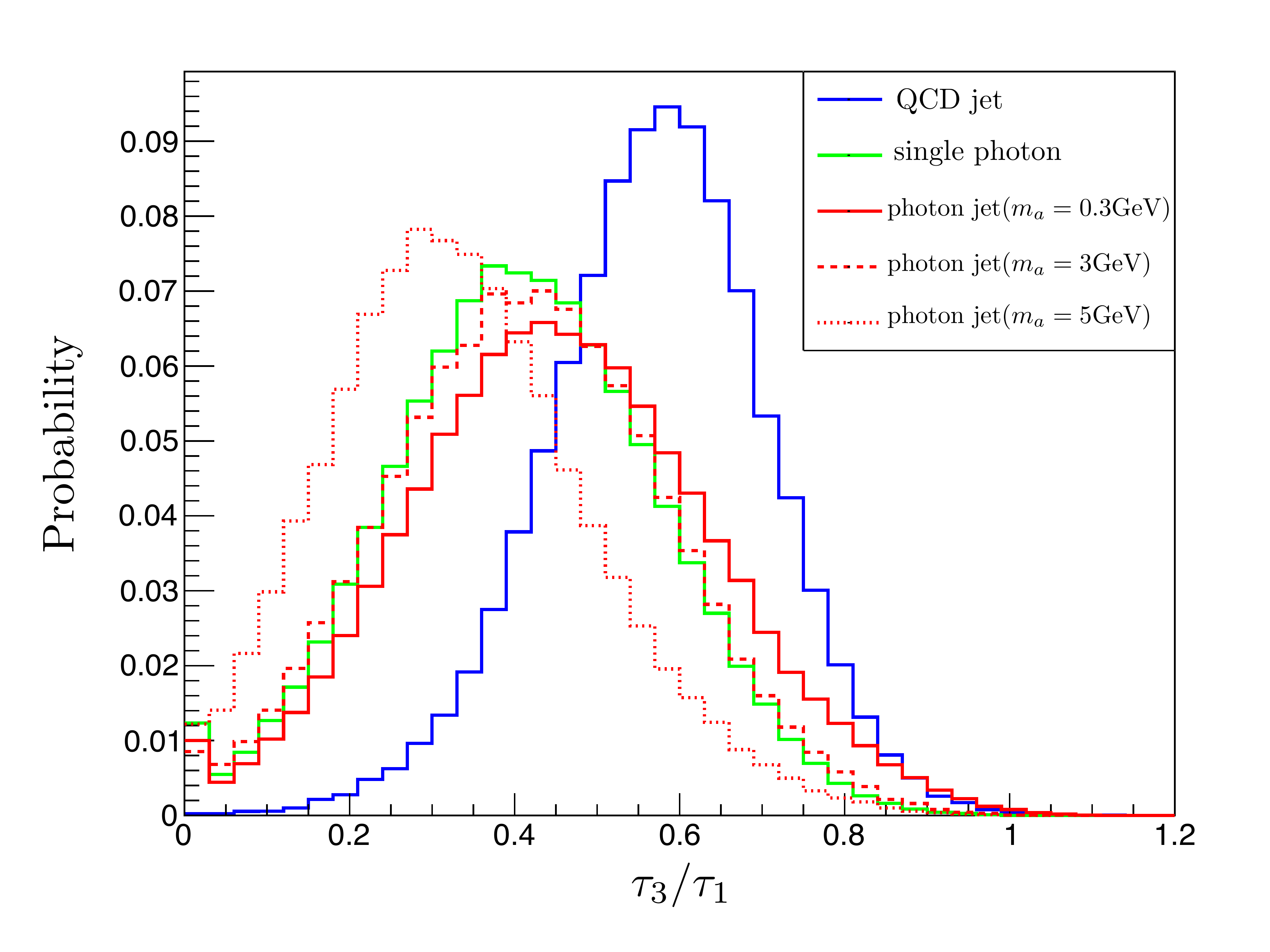}
\includegraphics[width=8.4cm]{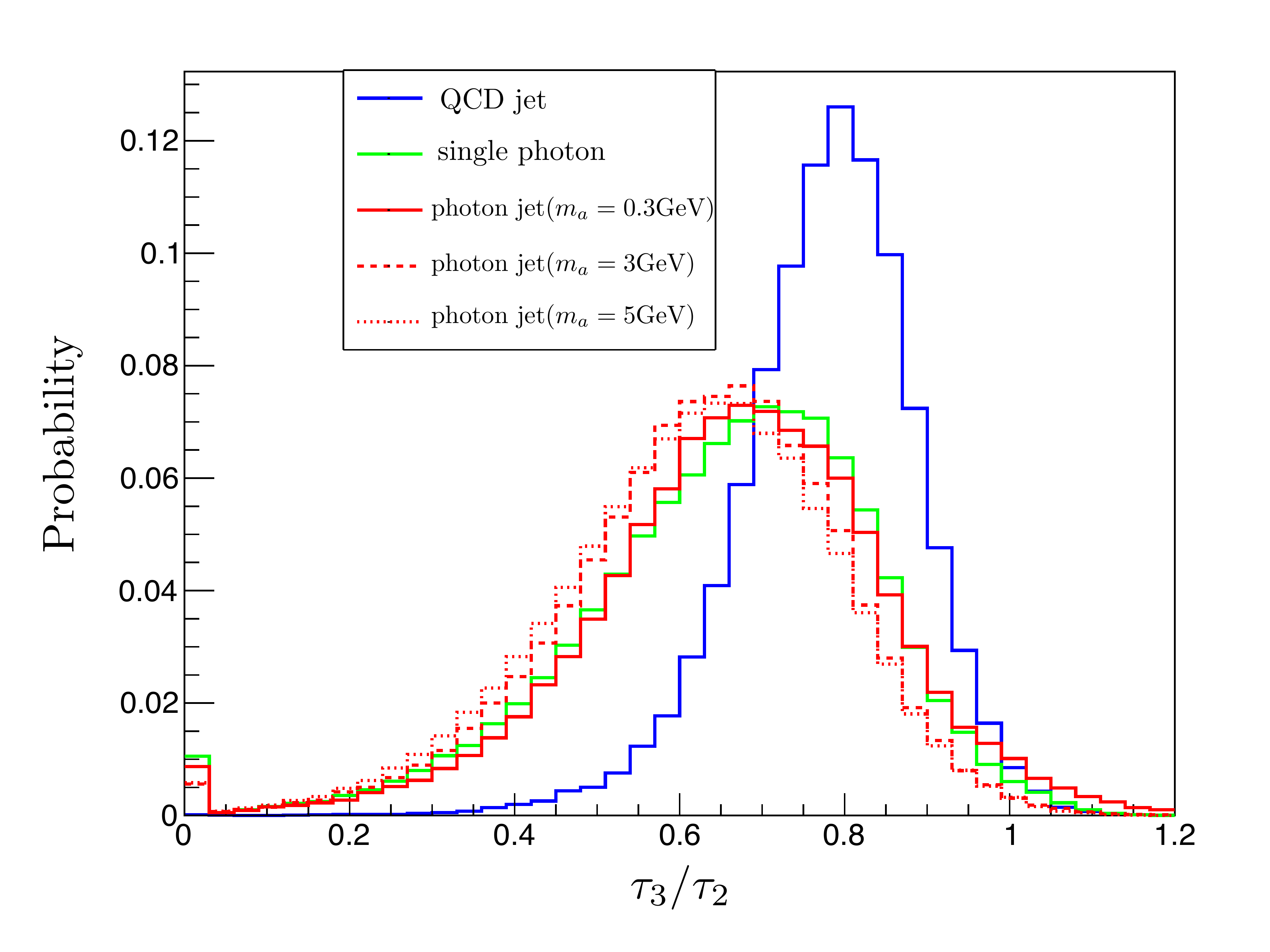}
\caption{Distributions of N-subjettiness variables (log$\tau_1$, $\tau_2/\tau_1$, $\tau_3/\tau_1$, $\tau_3/\tau_2$) for single photon, photon-jet and QCD jet events.}
\label{distributions2}
\end{center}
\end{figure*}
\begin{figure*}[ht]
\begin{center}
\includegraphics[width=8.4cm]{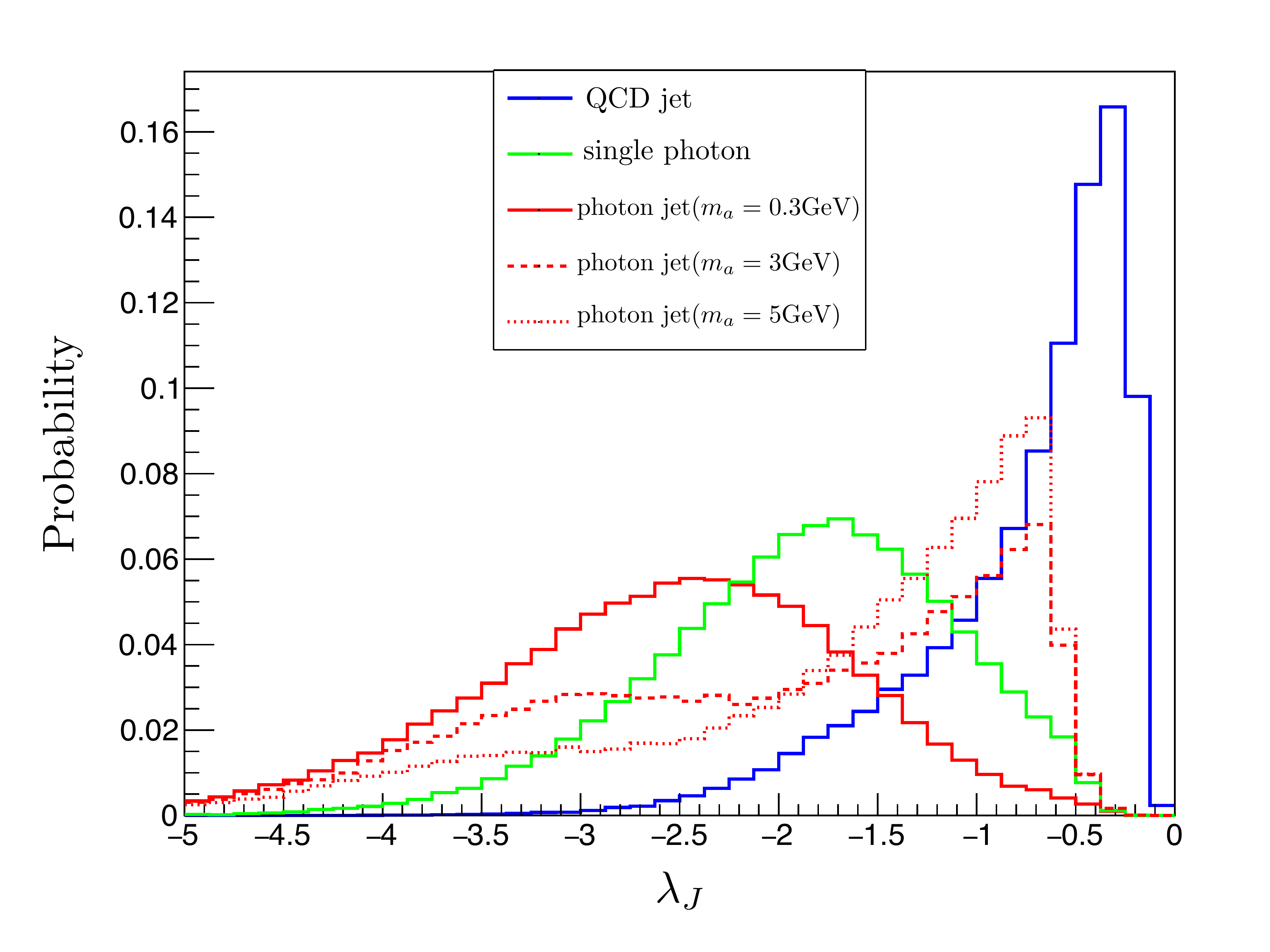}
\includegraphics[width=8.4cm]{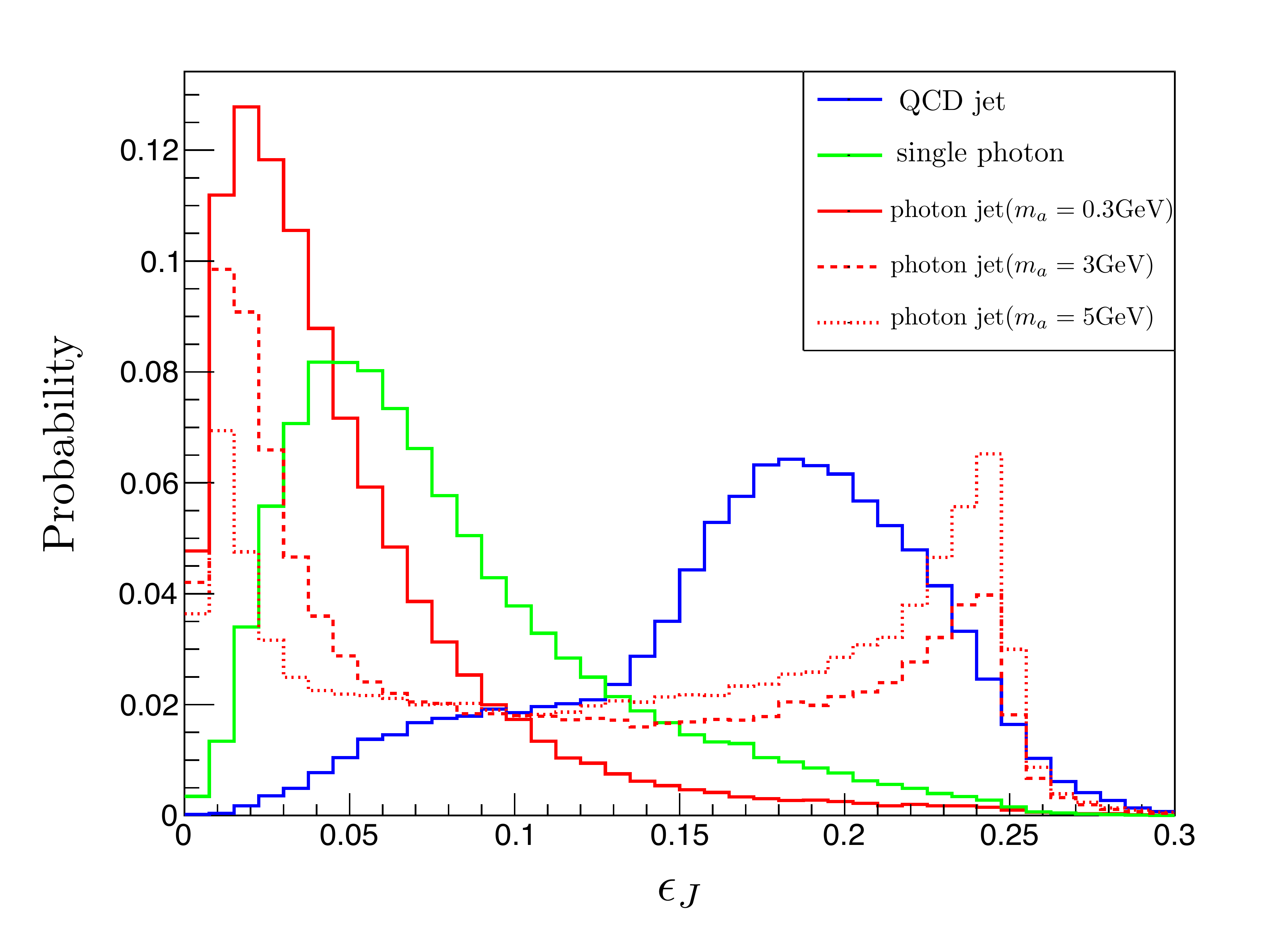}
\caption{Distributions of jet substructure variables ($\lambda_J$, $\epsilon_J$) for single photon, photon-jet, and QCD jet events.}
\label{distributions3}
\end{center}
\end{figure*}

\textit{Hadronic energy fraction of a jet}, $\theta_J$,  which is defined as the energy fraction carried by a jet's constituents that belong to the hadronic calorimeter (HCal), including the EFlowNeutralHadrons and ChargedHadrons in the energy-flow algorithm:
\begin{equation}
 \theta_J  \ = \ \frac{1} {E_J } (\sum_{i } E_{i} + \sum_{j} E_{j}),
\end{equation}
where $E_J$ is the total energy of the leading jet, $E_{i}$ and $E_{j}$ are the energy of the $i$-th EFlowNeutralHadron and the energy of the $j$-th ChargedHadron respectively that are constituents of the jet. Due to isospin symmetry, a QCD-jet typically contains about 2/3 charged pions and 1/3 neutral pions, which will decay to a pair of photons. Thus we expect to have a peak at $\theta_{J}\sim$ 2/3 (log$\theta_{J}\sim$ -0.2) for QCD jets. On the other hand, most of the energy of single photons and photon-jets are deposited in Electromagnetic Calorimeter (ECal). So the log$\theta_{J}$ of them are much smaller than that of QCD jets. In Fig.~\ref{distributions1}, we show the distribution of log$\theta_J$ for QCD jets, single photons, and photon-jets in our simulation data.

\textit{The number of charged tracks in a jet}, $N_{track}$,  which means the number of charged particles inside a jet. We calculate the angular distance $\Delta R$ between the leading jet and all the tracks with $p_T$ > 2 GeV. When $\Delta R < R_j$, the corresponding track is considered to be inside the leading jet.  As mentioned before, a QCD-jet typically contains several charged pions while single photons and photon-jets leave no tracks in the Tracker. Therefore, the number of tracks associated with QCD-jets varies over a broad range, however, the single photon and photon-jet samples are dominated by jets without associated tracks. In Fig.~\ref{distributions1}, we show the distribution of $N_{track}$ per jet for QCD jet, single photon and photon-jet samples. It can be seen that the variable $N_{track}$ has a good discrimination power of QCD-jets and single photons/photon-jets, but cannot distinguish the single photons from photon-jets.

\textit{N-subjettiness}, which is a measure of the number of energetic subjets inside a jet~\cite{Thaler:2010tr,Thaler:2011gf}. Given a set of N-axes, we can define
\begin{equation}
	\label{eq:tau}
	\tau_N = \frac{ \sum_k  p_{T_{k}}  \times \text{min} \bigl \{ \Delta R_{1,k}, \cdots, \Delta R_{N,k}  \bigr \} } { \sum_k  p_{T_{k}} \times R_j}  \; ,
\end{equation}
where $k$ runs over all the constituents of the jet. $p_{T_{k}}$ is the transverse momentum of the $k$-th constituent, 
and $\Delta  R_{l,k}$ is the angular distance between the $l$-th subjet and the $k$-th constituent of the jet. When $N=1$, N-subjettiness describes the energy distribution of the jet. In Fig.~\ref{distributions2}, it can be seen that the single photon and photon-jet samples peak at lower values of $\log \tau_1$, as a comparison with the QCD jets. Besides, $\tau_{N}$ will decrease rapidly as $N$ increasing. Thus, the ratio of two N-subjettiness, $\tau_a/\tau_b$, can be used to separate the signal and backgrounds. From Fig.~\ref{distributions2}, we can see that the photon-jet and single photon events have smaller values of $\tau_{2}/ \tau_{1}$, $\tau_{3}/ \tau_{1}$ and $\tau_{3}/ \tau_{2}$ than the QCD-jets.


\textit{Functions of energy and $p_{T}$ of subjets} We use the Jet Filtering~\cite{Butterworth:2008iy} in \textsf{FastJet} to recluster the leading jet's constituents using $k_T$ algorithm. After reclustering, we obtain $N$ exclusive $k_T$-subjets. We set $N=5$ and take the three hardest subjets to construct two jet substructure variables. One is the fraction of the jet $p_T$ carried by the leading subjet $\lambda_J$,
\begin{equation}
	\label{eq:lambda}
	\lambda_J \ = \ \log \Bigl(1- \frac{p_{T_{L}}} {p_{T_{J}}} \Bigr) \; ,
\end{equation}
and the other is the energy correlation function of the three hardest subjets $\epsilon_J$,
\begin{equation}
	\label{eq:epsilon}
 	\epsilon_J \ = \  \frac{1}{E_J^2}  \sum_{(i > j) \in N_\text{hard} }  E_{i} E_{j}\,,
\end{equation}
where $p_{T_{L}}$ and $E_J$ are the total momentum and energy of a given jet, respectively. $E_{i}$ is the energy of the $i$-th subjet. For the single photon and photon-jet, the leading subjet carries most of the energy. Therefore, their distributions of $\lambda_J$ have smaller values than QCD-jets. As for $\epsilon_J$, among the 3 hardest jets, $\epsilon_J$ increases with the number of subjets with similar energy. The photon-jet and QCD-jets have a larger values of  $\epsilon_J$ than the single photon. Fig.~\ref{distributions3} presents the corresponding distributions of $\epsilon_J$ and $\lambda_J$ for signal and backgrounds.

In the following we implement the above variables in a Boosted Decision Tree (BDT)~\cite{Roe:2004na} to enhance the ability of distinguishing signal from background, as Ref.~\cite{Ellis:2012sd,Ellis:2012zp}.
In practice, we use the Toolkit for Multivariate Analysis (TMVA)~\cite{Hocker:2007ht} package and the ``BDTD'' option to book BDTs. We uses 200 trees ensemble that requires a minimum training events in each node of $2.5\%$ and a maximum tree depth of 3. Other variables are set at their default values. It is trained using the half of the signal and background events and is tested on the rest of the events. We also demand the Kolmogorov-Smirnov test of the BDT analysis to be greater than 0.01 to avoid overtraining.

From Fig.~\ref{distributions1}, we can see that $\theta_J$ and $N_{track}$ offer a good discrimination between photon-jets and QCD jets. The other six variables can be used for discrimination between photon-jets and single photons. In order to achieve simultaneous separation of photon-jets, single photons, and QCD-jets, we take the $\theta_J$ and $N_{track}$ of photon-jets as signal samples and the $\theta_J$ and $N_{track}$ of QCD-jets as background samples to train the first BDT, namely BDT-1. The other six jet substructure variables of photon-jets and single photons, as signal samples and background samples respectively, are used to train the second BDT, namely BDT-2.
In other words, we discriminate photon-jet and single photon from QCD-jet by using BDT-1,
and then discriminate photon-jet from single photon by using BDT-2.

After training, BDT can map an event with two sets of variables \{log$\theta_J$, $N_{T}$\} and \{$\log \tau_1$, $\tau_{2}/ \tau_{1}$, $\tau_{3}/ \tau_{1}$, $\tau_{3}/ \tau_{2}$, $\lambda_J$, $\epsilon_J$\} into two BDT responses.

\begin{figure}[ht]
\begin{center}
\includegraphics[width=8cm]{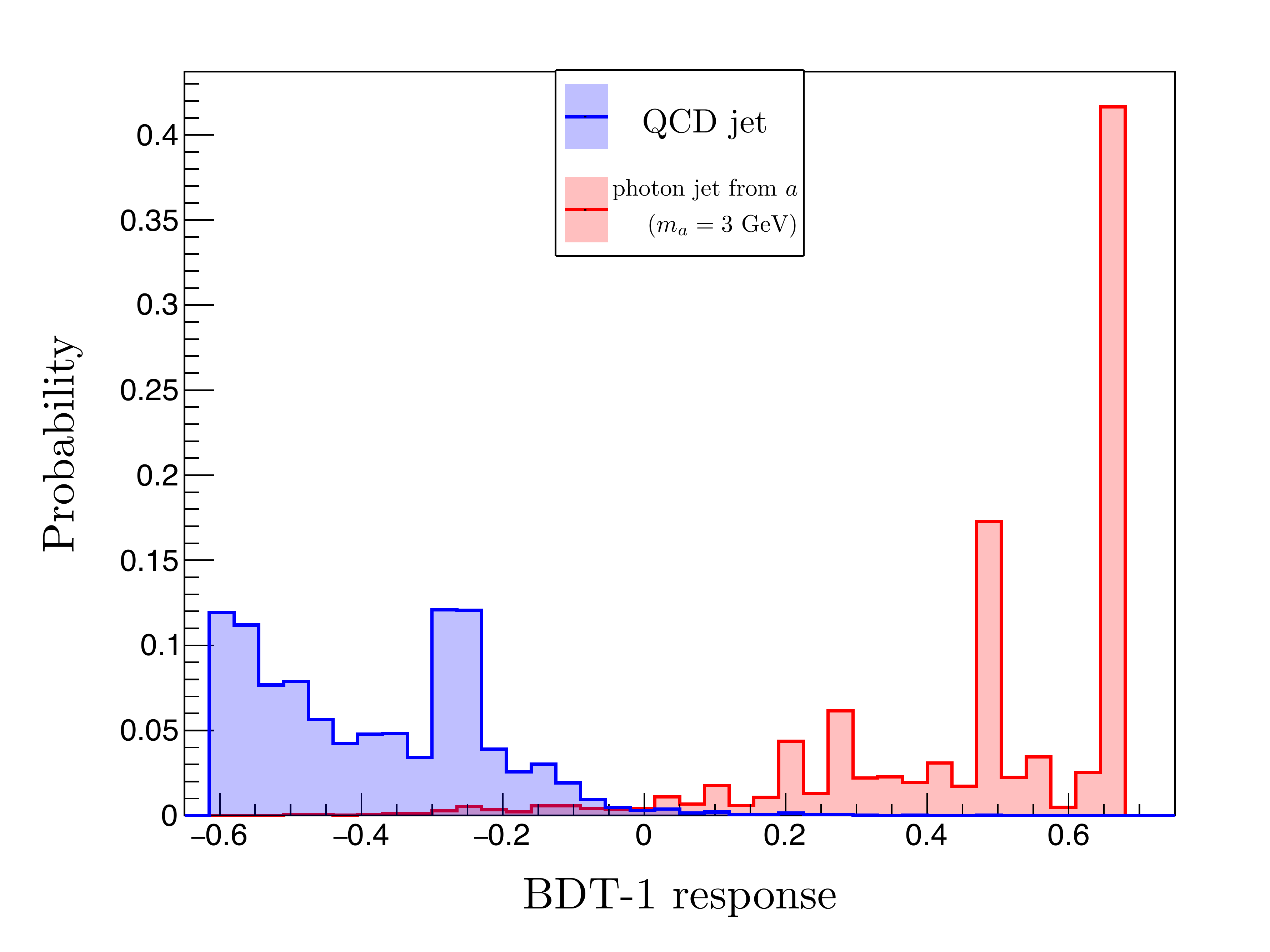}
\vspace{1mm}
\includegraphics[width=8cm]{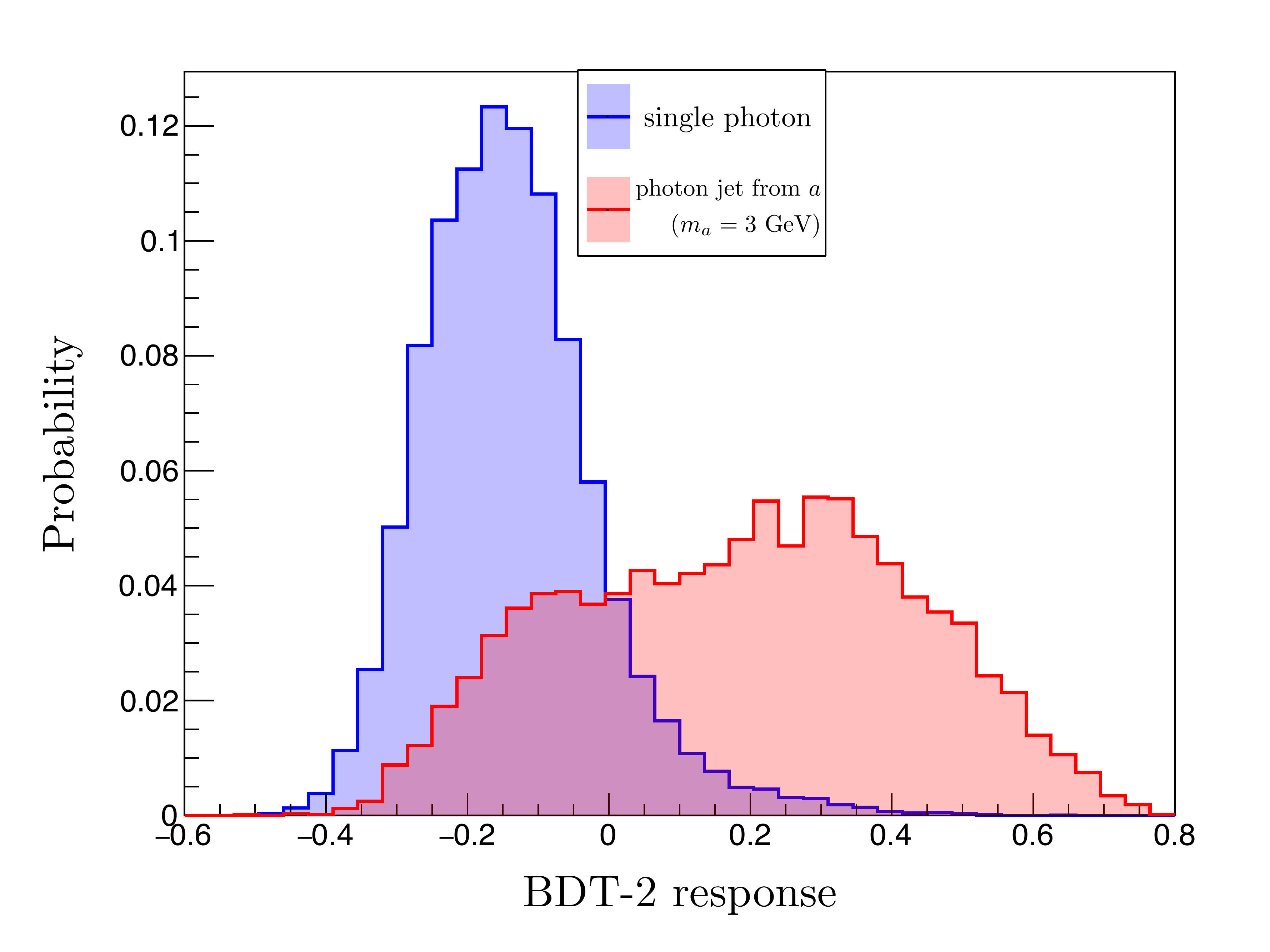}
\vspace{-2mm}
\caption{BDT response distributions of photon-jet as signal and single photon and QCD-jet as backgrounds respectively}
\label{BDT}
\end{center}
\end{figure}
In Fig.~\Ref{BDT} we show the BDT responses for $m_{a}$ = 3 GeV. In the BDT-1 analysis,  photon-jet-like events and photon-like events tend to get high BDT response while a QCD jet-like events tend to get low response. In the BDT-2, a photon-jet-like event tends to get high BDT response while a photon-like event tends to get a low BDT response. In BDT-1, we identify a jet with BDT response larger than 0 as a photon-jet or a single photon, while in BDT-2, we tag a jet with BDT response larger than 0.4 as a photon-jet. Since the jet substructure variables are sensitive to the ALP mass, we choose different BDT response cuts for different ALP masses to optimize the search ability.

\section{ALP-strahlung process at the LHC}
For the ALP-strahlung process $p p \to W^\pm a$, the final states are characterized by an isolated lepton and one photon-jet. We consider the SM backgrounds: QCD di-jet, $W^\pm j$, $W^\pm \gamma$, $t\bar{t}$ and $tj$. Based on the above analysis, we impose the following cuts to discriminate the signal and backgrounds: (i) Exactly one isolated lepton (electron or muon) with $p_\text{T} > 20$ GeV and $|\eta| < 2.5$; (ii) The hardest jet with $p_\text{T} > 50$ GeV and $|\eta| < 2.5$; (iii) The hardest jet's BDT-1 response is larger than 0 and BDT-2 response is larger than the corresponding BDT response cut for different ALP masses.

For the ALP-strahlung process $p p \to Z a$, the final states are characterized by the opposite sign and same flavor charged lepton pair and a photon-jet. The main SM backgrounds include $Z\gamma$ and $Zj$. According to our analysis, we discriminate the signal and backgrounds by imposing the following cuts:
(i) Exactly two leptons with $p_\text{T} > 20$ GeV and $|\eta| < 2.5$;
(ii) The invariant mass of the oppositely charged lepton pair with same flavor must be within $|m_{ll}-m_{Z}|$<20 GeV;
(iii) The hardest jet with $p_\text{T} > 50$ GeV and $|\eta| < 2.5$;
(iv) The hardest jet's BDT-1 response is larger than 0 and BDT-2 response is larger than the corresponding BDT response cut for each ALP mass.

\begin{table}[ht]
\begin{center}\begin{tabular}{|c|c|c|c|c|c|c|c|c|c|c|}
\hline  cut flow & \tabincell{c}{signal} &  \tabincell{c}{$jj$} & \tabincell{c}{$W^\pm\gamma$} & \tabincell{c}{ $W^\pm j$}  &\tabincell{c}{ $t\bar{t}$} & \tabincell{c}{$tj$}   \\
\hline  \tabincell{c}{1 lepton with \\$p_\text{T} > 20$ GeV \\and $|\eta| < 2.5$} & 36.24 & 19357.15 & 12.31 & 4448.53 & 151.86 & 29.39 \\
\hline  \tabincell{c}{The hardest jet \\with $p_\text{T} > 50$ GeV \\and $|\eta| < 2.5$} &23.31 & 12893 & 2.54 & 1605.92 & 136.41 & 18.90 \\
\hline  \tabincell{c}{The hardest jet's \\BDT-1 > 0 \\and BDT-2  > 0.4} &4.14 & 3.30 & 0.001 & 0.77 & 0.094 & 0.012 \\
\hline \end{tabular} \caption{The cut-flow of the cross sections (in units of pb) of the ALP-strahlung production process $pp \to aW^\pm$ and the corresponding backgrounds  at the 14 TeV LHC. The benchmark point is chosen as $m_{a}$=3 GeV and $g_{a\gamma\gamma}=16$ TeV$^{-1}$. }
\label{Cutflow6}
\end{center}
\end{table}
\begin{table}[ht]
\begin{center}\begin{tabular}{|c|c|c|c|c|c|c|c|c|c|c|}
\hline  ~~cut flow~~ & \tabincell{c}~~signal~~ &  \tabincell{c}{$Z\gamma$} & \tabincell{c}{$Zj$}  \\
\hline \tabincell{c}{2leptons with \\$p_\text{T} > 20$ GeV \\and $|\eta| < 2.5$} & 2.65 & 1.93 & 279.90 \\
\hline \tabincell{c}{oppositely charged lepton\\ pair with same flavor\\and  $|m_{ll}-m_{Z}|$<20 GeV} & 2.55 & 1.85 & 275.08 \\
\hline \tabincell{c}{The hardest jet \\with $p_\text{T} > 50$ GeV \\and $|\eta| < 2.5$} & 1.79 & 0.44 &~~ 104.476 ~~\\
\hline \tabincell{c}{The hardest jet's \\BDT-1 > 0 \\and BDT-2  > 0.4} & 0.29 &~~ 0.0003 ~~& 0.027 \\
\hline \end{tabular} \caption{Same as Table I, but for the ALP-strahlung production process $pp \to aZ$ and the corresponding backgrounds.}
\label{Cutflow5}
\end{center}
\end{table}

As an example, we consider a benchmark signal point with $m_{a}= 3$ GeV and $g_{a\gamma\gamma}=16$ TeV$^{-1}$. The cut-flows of this benchmark point and the main SM backgrounds are shown in Tables~\ref{Cutflow6} and \ref{Cutflow5}. It can be seen that after the BDT response cut, all the backgrounds are reduced dramatically. At the end of the cut flow, the largest backgrounds for the processes $p p \to W^\pm a$ and $p p \to Z a$ signal are QCD di-jet and $Zj$, respectively.

\begin{figure}[ht]
\begin{center}
\includegraphics[width=8cm]{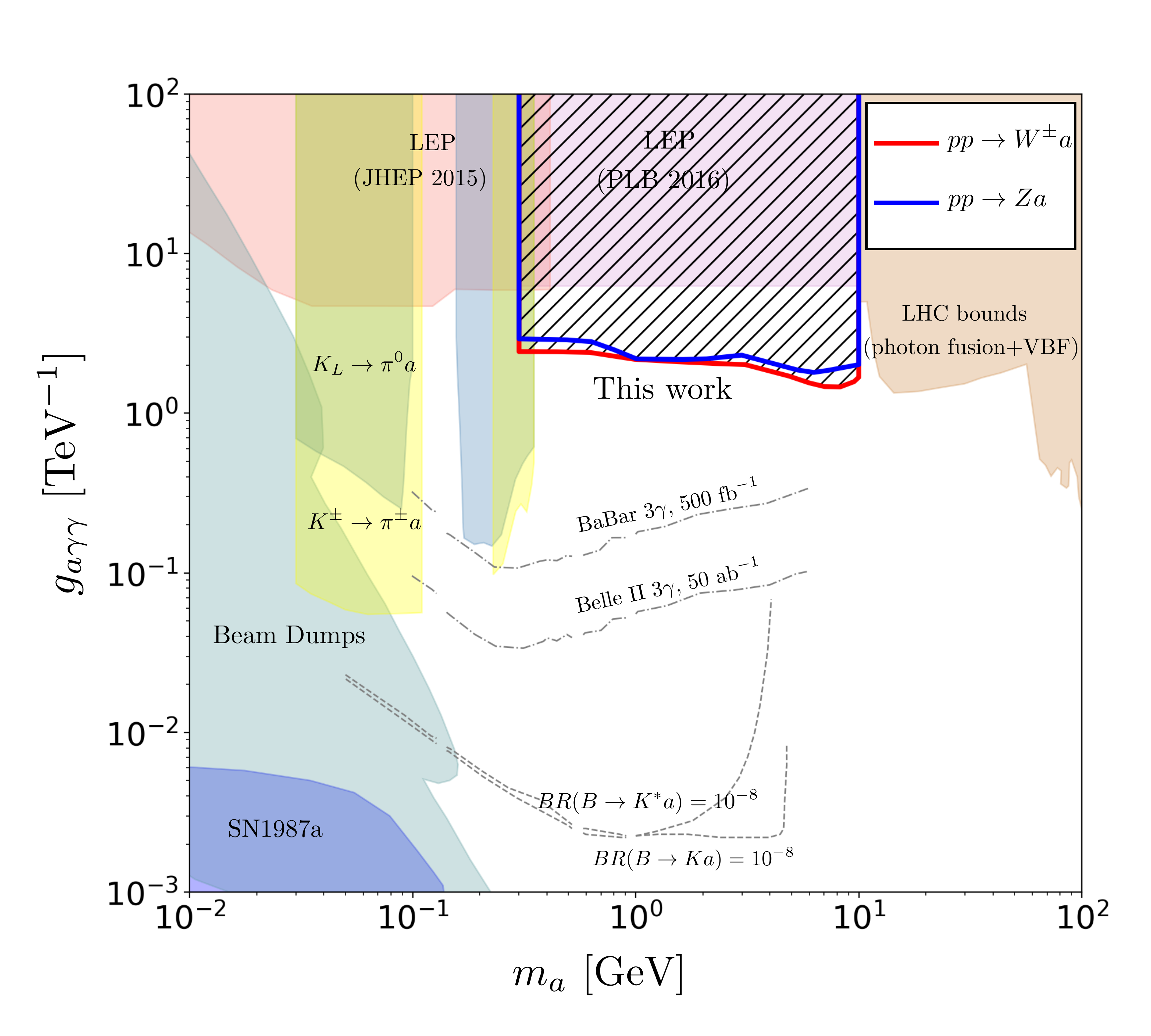}
\caption{The $2\sigma$ exclusion limits on the plane of the ALP mass $m_a$ versus the coupling $g_{a\gamma\gamma}$ from the ALP-strahlung production process $pp \to aW^\pm/Z$ at the LHC with luminosity 3000 fb$^{-1}$. Other limits from the LEP~\cite{Mimasu:2014nea,Jaeckel:2015jla}, the SN1987a~\cite{Payez:2014xsa,Jaeckel:2017tud}, the Beam Dump~\cite{Dobrich:2015jyk, Dobrich:2019dxc}, the ATLAS/CMS~\cite{ATLAS:2012fgo,Jaeckel:2012yz,Jaeckel:2015jla,Knapen:2016moh}, the rare $K$ and $B$ meson decays~\cite{Izaguirre:2016dfi} are also shown. It should be noted that the limits involving the ALP-gluon couplings are not shown but discussed in the context.}
\label{cm}
\end{center}
\end{figure}

In Fig.~\Ref{cm}, we present the $2\sigma$ bound from our ALP-strahlung production process $pp \to aW^\pm/Z$ on the plane of the ALP coupling $g_{a\gamma\gamma}$ versus the ALP mass $m_{a}$. We use the Poisson formula $\sqrt{2\left[(\mathcal{S}+\mathcal{B}){\rm ln}(1+\mathcal{S}/\mathcal{B})-\mathcal{S}\right]}$ to estimate the sensitivity at the LHC with the luminosity of 3000 fb$^{-1}$, where $\mathcal{S}$ and $\mathcal{B}$ are the number of signal and background events. As the comparison with our results, we also show other relevant bounds on the ALP-photon coupling. Since we only focus on the ALP-EW boson couplings in this work, the limits involving the ALP-gluon couplings are not shown in Fig.~\Ref{cm}.

It can be seen that our photon-jet method can cover the ALP mass region of 0.3 GeV $<m_a<$ 10 GeV at the LHC, which can extend the current LHC sensitivity of searching for the diphoton resonances in photon fusion and vector boson fusion processes~\cite{ATLAS:2012fgo,Jaeckel:2012yz,Jaeckel:2015jla,Knapen:2016moh}. In addition, our bound on the ALP-photon coupling $g_{a\gamma\gamma}$ is stronger than that derived from the null results of searching for the ALPs in $e^+ e^- \to 2\gamma/3\gamma$ processes  at the LEP~\cite{Jaeckel:2015jla}. For example, $g_{a\gamma\gamma}$ in our study is constrained to be less than 2.4 TeV$^{-1}$ at $m_a=0.3$ GeV and 1.5 TeV$^{-1}$ at $m_a=10$ GeV. On the other hand, we note that searching for $3\gamma$ events from the process $e^+e^- \to a(\to \gamma\gamma)\gamma$ and the rare meson decays, such as $K_L \to \pi^0 a$, $K^\pm \to \pi^\pm a$ and $B \to K^{(*)}a$, in the existing and future low energy experiments may provide more stringent constraints than ours when $ m_a \lesssim 5$ GeV~\cite{Izaguirre:2016dfi}.

Besides, if the ALP-gluon couplings are present and not suppressed by at
least two orders of magnitude relative to the electroweak ones, the current and future measurements of $B$ meson decays at the LHCb and B-factories can also give strong exclusion limits for $m_a<10$ GeV~\cite{CidVidal:2018blh,BaBar:2011kau}. However, it should be mentioned that this result is not applicable to our case because the ALP-gluon couplings are absent in our scenario. For the same reason, we do not show the constraints from isolated and energetic photons from hadronic decay of $Z$ boson at L3~\cite{Adriani:1992zm} and the searches of ATLAS and CMS for $\gamma\gamma$ resonances in the gluon fusion process~\cite{Mariotti:2017vtv}.


\section{Conclusions}
In this work we extended the current LHC sensitivity of probing the electroweak ALP via the ALP-strahlung production processes $pp \to aW^{\pm}/Z$ in the mass range of 0.3 GeV $<m_a<$ 10 GeV at the 14 TeV HL-LHC. Since the two photons from ALP decay are not well separated for such a light ALP, we used jet substructure variables and proposed photon-jet method to discriminate our signal from QCD jets and single photons backgrounds. With the help of BDT technique, we obtained the $2\sigma$ bounds on the ALP-photon coupling $g_{a\gamma\gamma}$ as a function of $m_{a}$. The coupling $g_{a\gamma\gamma}>$ 2.4 TeV$^{-1}$ at $m_a=0.3$ GeV and $g_{a\gamma\gamma}>$ 1.5 TeV$^{-1}$ at $m_a=10$ GeV can be excluded at $2\sigma$ level at the 14 TeV HL-LHC. This demonstrates that our approach can not only cover the mass gap 0.3 GeV $<m_a<$ 10 GeV at the LHC, but also surpass the existing LEP bounds. Although the measurements of rare meson decays may also exclude the ALP in MeV-scale mass range, a direct LHC search would provide an independent probe of this parameter space.

\section{acknowledgments}
This work is supported by the National Natural Science Foundation of China (NNSFC) under grant Nos. 11805161,
 12075300, 11821505, 11947118, by Peng-Huan-Wu Theoretical Physics Innovation Center (12047503), by the CAS Center for Excellence in Particle Physics (CCEPP), by the CAS Key
Research Program of Frontier Sciences, by a Key R\&D Program of Ministry of Science and Technology under number 2017YFA0402204, and by the Key Research Program of the Chinese Academy of Sciences, grant No. XDPB15.

\bibliography{ref}

\begin{thebibliography}{65}
\expandafter\ifx\csname natexlab\endcsname\relax\def\natexlab#1{#1}\fi
\expandafter\ifx\csname bibnamefont\endcsname\relax
  \def\bibnamefont#1{#1}\fi
\expandafter\ifx\csname bibfnamefont\endcsname\relax
  \def\bibfnamefont#1{#1}\fi
\expandafter\ifx\csname citenamefont\endcsname\relax
  \def\citenamefont#1{#1}\fi
\expandafter\ifx\csname url\endcsname\relax
  \def\url#1{\texttt{#1}}\fi
\expandafter\ifx\csname urlprefix\endcsname\relax\def\urlprefix{URL }\fi
\providecommand{\bibinfo}[2]{#2}
\providecommand{\eprint}[2][]{\url{#2}}

\bibitem[{\citenamefont{Peccei and Quinn}(1977)}]{Peccei:1977hh}
\bibinfo{author}{\bibfnamefont{R.~D.} \bibnamefont{Peccei}} \bibnamefont{and}
  \bibinfo{author}{\bibfnamefont{H.~R.} \bibnamefont{Quinn}},
  \bibinfo{journal}{Phys. Rev. Lett.} \textbf{\bibinfo{volume}{38}},
  \bibinfo{pages}{1440} (\bibinfo{year}{1977}).

\bibitem[{\citenamefont{Weinberg}(1978)}]{Weinberg:1977ma}
\bibinfo{author}{\bibfnamefont{S.}~\bibnamefont{Weinberg}},
  \bibinfo{journal}{Phys. Rev. Lett.} \textbf{\bibinfo{volume}{40}},
  \bibinfo{pages}{223} (\bibinfo{year}{1978}).

\bibitem[{\citenamefont{Wilczek}(1978)}]{Wilczek:1977pj}
\bibinfo{author}{\bibfnamefont{F.}~\bibnamefont{Wilczek}},
  \bibinfo{journal}{Phys. Rev. Lett.} \textbf{\bibinfo{volume}{40}},
  \bibinfo{pages}{279} (\bibinfo{year}{1978}).

\bibitem[{\citenamefont{Kim}(1979)}]{Kim:1979if}
\bibinfo{author}{\bibfnamefont{J.~E.} \bibnamefont{Kim}},
  \bibinfo{journal}{Phys. Rev. Lett.} \textbf{\bibinfo{volume}{43}},
  \bibinfo{pages}{103} (\bibinfo{year}{1979}).

\bibitem[{\citenamefont{Svrcek and Witten}(2006)}]{Svrcek:2006yi}
\bibinfo{author}{\bibfnamefont{P.}~\bibnamefont{Svrcek}} \bibnamefont{and}
  \bibinfo{author}{\bibfnamefont{E.}~\bibnamefont{Witten}},
  \bibinfo{journal}{JHEP} \textbf{\bibinfo{volume}{06}}, \bibinfo{pages}{051}
  (\bibinfo{year}{2006}), \eprint{hep-th/0605206}.

\bibitem[{\citenamefont{Arvanitaki et~al.}(2010)\citenamefont{Arvanitaki,
  Dimopoulos, Dubovsky, Kaloper, and March-Russell}}]{Arvanitaki:2009fg}
\bibinfo{author}{\bibfnamefont{A.}~\bibnamefont{Arvanitaki}},
  \bibinfo{author}{\bibfnamefont{S.}~\bibnamefont{Dimopoulos}},
  \bibinfo{author}{\bibfnamefont{S.}~\bibnamefont{Dubovsky}},
  \bibinfo{author}{\bibfnamefont{N.}~\bibnamefont{Kaloper}}, \bibnamefont{and}
  \bibinfo{author}{\bibfnamefont{J.}~\bibnamefont{March-Russell}},
  \bibinfo{journal}{Phys. Rev. D} \textbf{\bibinfo{volume}{81}},
  \bibinfo{pages}{123530} (\bibinfo{year}{2010}), \eprint{0905.4720}.

\bibitem[{\citenamefont{Cicoli et~al.}(2012)\citenamefont{Cicoli, Goodsell, and
  Ringwald}}]{Cicoli:2012sz}
\bibinfo{author}{\bibfnamefont{M.}~\bibnamefont{Cicoli}},
  \bibinfo{author}{\bibfnamefont{M.}~\bibnamefont{Goodsell}}, \bibnamefont{and}
  \bibinfo{author}{\bibfnamefont{A.}~\bibnamefont{Ringwald}},
  \bibinfo{journal}{JHEP} \textbf{\bibinfo{volume}{10}}, \bibinfo{pages}{146}
  (\bibinfo{year}{2012}), \eprint{1206.0819}.

\bibitem[{\citenamefont{Ballesteros et~al.}(2017)\citenamefont{Ballesteros,
  Redondo, Ringwald, and Tamarit}}]{Ballesteros:2016euj}
\bibinfo{author}{\bibfnamefont{G.}~\bibnamefont{Ballesteros}},
  \bibinfo{author}{\bibfnamefont{J.}~\bibnamefont{Redondo}},
  \bibinfo{author}{\bibfnamefont{A.}~\bibnamefont{Ringwald}}, \bibnamefont{and}
  \bibinfo{author}{\bibfnamefont{C.}~\bibnamefont{Tamarit}},
  \bibinfo{journal}{Phys. Rev. Lett.} \textbf{\bibinfo{volume}{118}},
  \bibinfo{pages}{071802} (\bibinfo{year}{2017}), \eprint{1608.05414}.

\bibitem[{\citenamefont{Graham et~al.}(2015)\citenamefont{Graham, Kaplan, and
  Rajendran}}]{Graham:2015cka}
\bibinfo{author}{\bibfnamefont{P.~W.} \bibnamefont{Graham}},
  \bibinfo{author}{\bibfnamefont{D.~E.} \bibnamefont{Kaplan}},
  \bibnamefont{and}
  \bibinfo{author}{\bibfnamefont{S.}~\bibnamefont{Rajendran}},
  \bibinfo{journal}{Phys. Rev. Lett.} \textbf{\bibinfo{volume}{115}},
  \bibinfo{pages}{221801} (\bibinfo{year}{2015}), \eprint{1504.07551}.

\bibitem[{\citenamefont{Raffelt}(1990)}]{Raffelt:1990yz}
\bibinfo{author}{\bibfnamefont{G.~G.} \bibnamefont{Raffelt}},
  \bibinfo{journal}{Phys. Rept.} \textbf{\bibinfo{volume}{198}},
  \bibinfo{pages}{1} (\bibinfo{year}{1990}).

\bibitem[{\citenamefont{Marsh}(2016)}]{Marsh:2015xka}
\bibinfo{author}{\bibfnamefont{D.~J.~E.} \bibnamefont{Marsh}},
  \bibinfo{journal}{Phys. Rept.} \textbf{\bibinfo{volume}{643}},
  \bibinfo{pages}{1} (\bibinfo{year}{2016}), \eprint{1510.07633}.

\bibitem[{\citenamefont{Preskill et~al.}(1983)\citenamefont{Preskill, Wise, and
  Wilczek}}]{Preskill:1982cy}
\bibinfo{author}{\bibfnamefont{J.}~\bibnamefont{Preskill}},
  \bibinfo{author}{\bibfnamefont{M.~B.} \bibnamefont{Wise}}, \bibnamefont{and}
  \bibinfo{author}{\bibfnamefont{F.}~\bibnamefont{Wilczek}},
  \bibinfo{journal}{Phys. Lett. B} \textbf{\bibinfo{volume}{120}},
  \bibinfo{pages}{127} (\bibinfo{year}{1983}).

\bibitem[{\citenamefont{Abbott and Sikivie}(1983)}]{Abbott:1982af}
\bibinfo{author}{\bibfnamefont{L.~F.} \bibnamefont{Abbott}} \bibnamefont{and}
  \bibinfo{author}{\bibfnamefont{P.}~\bibnamefont{Sikivie}},
  \bibinfo{journal}{Phys. Lett. B} \textbf{\bibinfo{volume}{120}},
  \bibinfo{pages}{133} (\bibinfo{year}{1983}).

\bibitem[{\citenamefont{Dine and Fischler}(1983)}]{Dine:1982ah}
\bibinfo{author}{\bibfnamefont{M.}~\bibnamefont{Dine}} \bibnamefont{and}
  \bibinfo{author}{\bibfnamefont{W.}~\bibnamefont{Fischler}},
  \bibinfo{journal}{Phys. Lett. B} \textbf{\bibinfo{volume}{120}},
  \bibinfo{pages}{137} (\bibinfo{year}{1983}).

\bibitem[{\citenamefont{Arias et~al.}(2012)\citenamefont{Arias, Cadamuro,
  Goodsell, Jaeckel, Redondo, and Ringwald}}]{Arias:2012az}
\bibinfo{author}{\bibfnamefont{P.}~\bibnamefont{Arias}},
  \bibinfo{author}{\bibfnamefont{D.}~\bibnamefont{Cadamuro}},
  \bibinfo{author}{\bibfnamefont{M.}~\bibnamefont{Goodsell}},
  \bibinfo{author}{\bibfnamefont{J.}~\bibnamefont{Jaeckel}},
  \bibinfo{author}{\bibfnamefont{J.}~\bibnamefont{Redondo}}, \bibnamefont{and}
  \bibinfo{author}{\bibfnamefont{A.}~\bibnamefont{Ringwald}},
  \bibinfo{journal}{JCAP} \textbf{\bibinfo{volume}{06}}, \bibinfo{pages}{013}
  (\bibinfo{year}{2012}), \eprint{1201.5902}.

\bibitem[{\citenamefont{Jaeckel et~al.}(2014)\citenamefont{Jaeckel, Redondo,
  and Ringwald}}]{Jaeckel:2014qea}
\bibinfo{author}{\bibfnamefont{J.}~\bibnamefont{Jaeckel}},
  \bibinfo{author}{\bibfnamefont{J.}~\bibnamefont{Redondo}}, \bibnamefont{and}
  \bibinfo{author}{\bibfnamefont{A.}~\bibnamefont{Ringwald}},
  \bibinfo{journal}{Phys. Rev. D} \textbf{\bibinfo{volume}{89}},
  \bibinfo{pages}{103511} (\bibinfo{year}{2014}), \eprint{1402.7335}.

\bibitem[{\citenamefont{Athron et~al.}(2021)}]{Athron:2020maw}
\bibinfo{author}{\bibfnamefont{P.}~\bibnamefont{Athron}} \bibnamefont{et~al.},
  \bibinfo{journal}{JHEP} \textbf{\bibinfo{volume}{05}}, \bibinfo{pages}{159}
  (\bibinfo{year}{2021}), \eprint{2007.05517}.

\bibitem[{\citenamefont{Gao et~al.}(2020)\citenamefont{Gao, Liu, Wang, Wang,
  Xue, and Zhong}}]{Gao:2020wer}
\bibinfo{author}{\bibfnamefont{C.}~\bibnamefont{Gao}},
  \bibinfo{author}{\bibfnamefont{J.}~\bibnamefont{Liu}},
  \bibinfo{author}{\bibfnamefont{L.-T.} \bibnamefont{Wang}},
  \bibinfo{author}{\bibfnamefont{X.-P.} \bibnamefont{Wang}},
  \bibinfo{author}{\bibfnamefont{W.}~\bibnamefont{Xue}}, \bibnamefont{and}
  \bibinfo{author}{\bibfnamefont{Y.-M.} \bibnamefont{Zhong}},
  \bibinfo{journal}{Phys. Rev. Lett.} \textbf{\bibinfo{volume}{125}},
  \bibinfo{pages}{131806} (\bibinfo{year}{2020}), \eprint{2006.14598}.

\bibitem[{\citenamefont{Izaguirre et~al.}(2017)\citenamefont{Izaguirre, Lin,
  and Shuve}}]{Izaguirre:2016dfi}
\bibinfo{author}{\bibfnamefont{E.}~\bibnamefont{Izaguirre}},
  \bibinfo{author}{\bibfnamefont{T.}~\bibnamefont{Lin}}, \bibnamefont{and}
  \bibinfo{author}{\bibfnamefont{B.}~\bibnamefont{Shuve}},
  \bibinfo{journal}{Phys. Rev. Lett.} \textbf{\bibinfo{volume}{118}},
  \bibinfo{pages}{111802} (\bibinfo{year}{2017}), \eprint{1611.09355}.

\bibitem[{\citenamefont{Dolan et~al.}(2017)\citenamefont{Dolan, Ferber, Hearty,
  Kahlhoefer, and Schmidt-Hoberg}}]{Dolan:2017osp}
\bibinfo{author}{\bibfnamefont{M.~J.} \bibnamefont{Dolan}},
  \bibinfo{author}{\bibfnamefont{T.}~\bibnamefont{Ferber}},
  \bibinfo{author}{\bibfnamefont{C.}~\bibnamefont{Hearty}},
  \bibinfo{author}{\bibfnamefont{F.}~\bibnamefont{Kahlhoefer}},
  \bibnamefont{and}
  \bibinfo{author}{\bibfnamefont{K.}~\bibnamefont{Schmidt-Hoberg}},
  \bibinfo{journal}{JHEP} \textbf{\bibinfo{volume}{12}}, \bibinfo{pages}{094}
  (\bibinfo{year}{2017}), \eprint{1709.00009}.

\bibitem[{\citenamefont{Bauer et~al.}(2020)\citenamefont{Bauer, Neubert,
  Renner, Schnubel, and Thamm}}]{Bauer:2019gfk}
\bibinfo{author}{\bibfnamefont{M.}~\bibnamefont{Bauer}},
  \bibinfo{author}{\bibfnamefont{M.}~\bibnamefont{Neubert}},
  \bibinfo{author}{\bibfnamefont{S.}~\bibnamefont{Renner}},
  \bibinfo{author}{\bibfnamefont{M.}~\bibnamefont{Schnubel}}, \bibnamefont{and}
  \bibinfo{author}{\bibfnamefont{A.}~\bibnamefont{Thamm}},
  \bibinfo{journal}{Phys. Rev. Lett.} \textbf{\bibinfo{volume}{124}},
  \bibinfo{pages}{211803} (\bibinfo{year}{2020}), \eprint{1908.00008}.

\bibitem[{\citenamefont{Banerjee et~al.}(2020)}]{Banerjee:2020fue}
\bibinfo{author}{\bibfnamefont{D.}~\bibnamefont{Banerjee}} \bibnamefont{et~al.}
  (\bibinfo{collaboration}{NA64}), \bibinfo{journal}{Phys. Rev. Lett.}
  \textbf{\bibinfo{volume}{125}}, \bibinfo{pages}{081801}
  (\bibinfo{year}{2020}), \eprint{2005.02710}.

\bibitem[{\citenamefont{Gu et~al.}(2021)\citenamefont{Gu, Wu, and
  Zhu}}]{Gu:2021lni}
\bibinfo{author}{\bibfnamefont{Y.}~\bibnamefont{Gu}},
  \bibinfo{author}{\bibfnamefont{L.}~\bibnamefont{Wu}}, \bibnamefont{and}
  \bibinfo{author}{\bibfnamefont{B.}~\bibnamefont{Zhu}} (\bibinfo{year}{2021}),
  \eprint{2105.07232}.

\bibitem[{\citenamefont{Mimasu and Sanz}(2015)}]{Mimasu:2014nea}
\bibinfo{author}{\bibfnamefont{K.}~\bibnamefont{Mimasu}} \bibnamefont{and}
  \bibinfo{author}{\bibfnamefont{V.}~\bibnamefont{Sanz}},
  \bibinfo{journal}{JHEP} \textbf{\bibinfo{volume}{06}}, \bibinfo{pages}{173}
  (\bibinfo{year}{2015}), \eprint{1409.4792}.

\bibitem[{\citenamefont{Knapen et~al.}(2017)\citenamefont{Knapen, Lin, Lou, and
  Melia}}]{Knapen:2016moh}
\bibinfo{author}{\bibfnamefont{S.}~\bibnamefont{Knapen}},
  \bibinfo{author}{\bibfnamefont{T.}~\bibnamefont{Lin}},
  \bibinfo{author}{\bibfnamefont{H.~K.} \bibnamefont{Lou}}, \bibnamefont{and}
  \bibinfo{author}{\bibfnamefont{T.}~\bibnamefont{Melia}},
  \bibinfo{journal}{Phys. Rev. Lett.} \textbf{\bibinfo{volume}{118}},
  \bibinfo{pages}{171801} (\bibinfo{year}{2017}), \eprint{1607.06083}.

\bibitem[{\citenamefont{Barrie et~al.}(2016)\citenamefont{Barrie, Kobakhidze,
  Talia, and Wu}}]{Barrie:2016ntq}
\bibinfo{author}{\bibfnamefont{N.~D.} \bibnamefont{Barrie}},
  \bibinfo{author}{\bibfnamefont{A.}~\bibnamefont{Kobakhidze}},
  \bibinfo{author}{\bibfnamefont{M.}~\bibnamefont{Talia}}, \bibnamefont{and}
  \bibinfo{author}{\bibfnamefont{L.}~\bibnamefont{Wu}}, \bibinfo{journal}{Phys.
  Lett. B} \textbf{\bibinfo{volume}{755}}, \bibinfo{pages}{343}
  (\bibinfo{year}{2016}), \eprint{1602.00475}.

\bibitem[{\citenamefont{Bauer et~al.}(2017{\natexlab{a}})\citenamefont{Bauer,
  Neubert, and Thamm}}]{Bauer:2017ris}
\bibinfo{author}{\bibfnamefont{M.}~\bibnamefont{Bauer}},
  \bibinfo{author}{\bibfnamefont{M.}~\bibnamefont{Neubert}}, \bibnamefont{and}
  \bibinfo{author}{\bibfnamefont{A.}~\bibnamefont{Thamm}},
  \bibinfo{journal}{JHEP} \textbf{\bibinfo{volume}{12}}, \bibinfo{pages}{044}
  (\bibinfo{year}{2017}{\natexlab{a}}), \eprint{1708.00443}.

\bibitem[{\citenamefont{Brivio et~al.}(2017)\citenamefont{Brivio, Gavela,
  Merlo, Mimasu, No, del Rey, and Sanz}}]{Brivio:2017ije}
\bibinfo{author}{\bibfnamefont{I.}~\bibnamefont{Brivio}},
  \bibinfo{author}{\bibfnamefont{M.~B.} \bibnamefont{Gavela}},
  \bibinfo{author}{\bibfnamefont{L.}~\bibnamefont{Merlo}},
  \bibinfo{author}{\bibfnamefont{K.}~\bibnamefont{Mimasu}},
  \bibinfo{author}{\bibfnamefont{J.~M.} \bibnamefont{No}},
  \bibinfo{author}{\bibfnamefont{R.}~\bibnamefont{del Rey}}, \bibnamefont{and}
  \bibinfo{author}{\bibfnamefont{V.}~\bibnamefont{Sanz}},
  \bibinfo{journal}{Eur. Phys. J. C} \textbf{\bibinfo{volume}{77}},
  \bibinfo{pages}{572} (\bibinfo{year}{2017}), \eprint{1701.05379}.

\bibitem[{\citenamefont{Bauer et~al.}(2019)\citenamefont{Bauer, Heiles,
  Neubert, and Thamm}}]{Bauer:2018uxu}
\bibinfo{author}{\bibfnamefont{M.}~\bibnamefont{Bauer}},
  \bibinfo{author}{\bibfnamefont{M.}~\bibnamefont{Heiles}},
  \bibinfo{author}{\bibfnamefont{M.}~\bibnamefont{Neubert}}, \bibnamefont{and}
  \bibinfo{author}{\bibfnamefont{A.}~\bibnamefont{Thamm}},
  \bibinfo{journal}{Eur. Phys. J. C} \textbf{\bibinfo{volume}{79}},
  \bibinfo{pages}{74} (\bibinfo{year}{2019}), \eprint{1808.10323}.

\bibitem[{\citenamefont{Ebadi et~al.}(2019)\citenamefont{Ebadi, Khatibi, and
  Mohammadi~Najafabadi}}]{Ebadi:2019gij}
\bibinfo{author}{\bibfnamefont{J.}~\bibnamefont{Ebadi}},
  \bibinfo{author}{\bibfnamefont{S.}~\bibnamefont{Khatibi}}, \bibnamefont{and}
  \bibinfo{author}{\bibfnamefont{M.}~\bibnamefont{Mohammadi~Najafabadi}},
  \bibinfo{journal}{Phys. Rev. D} \textbf{\bibinfo{volume}{100}},
  \bibinfo{pages}{015016} (\bibinfo{year}{2019}), \eprint{1901.03061}.

\bibitem[{\citenamefont{Wang et~al.}(2021)\citenamefont{Wang, Wu, and
  Zhang}}]{Wang:2020ips}
\bibinfo{author}{\bibfnamefont{D.}~\bibnamefont{Wang}},
  \bibinfo{author}{\bibfnamefont{L.}~\bibnamefont{Wu}}, \bibnamefont{and}
  \bibinfo{author}{\bibfnamefont{M.}~\bibnamefont{Zhang}},
  \bibinfo{journal}{Phys. Rev. D} \textbf{\bibinfo{volume}{103}},
  \bibinfo{pages}{115017} (\bibinfo{year}{2021}), \eprint{2007.09722}.

\bibitem[{\citenamefont{Ren et~al.}(2021)\citenamefont{Ren, Wang, Wu, Yang, and
  Zhang}}]{Ren:2021prq}
\bibinfo{author}{\bibfnamefont{J.}~\bibnamefont{Ren}},
  \bibinfo{author}{\bibfnamefont{D.}~\bibnamefont{Wang}},
  \bibinfo{author}{\bibfnamefont{L.}~\bibnamefont{Wu}},
  \bibinfo{author}{\bibfnamefont{J.~M.} \bibnamefont{Yang}}, \bibnamefont{and}
  \bibinfo{author}{\bibfnamefont{M.}~\bibnamefont{Zhang}}
  (\bibinfo{year}{2021}), \eprint{2106.07018}.

\bibitem[{\citenamefont{Jaeckel and Spannowsky}(2016)}]{Jaeckel:2015jla}
\bibinfo{author}{\bibfnamefont{J.}~\bibnamefont{Jaeckel}} \bibnamefont{and}
  \bibinfo{author}{\bibfnamefont{M.}~\bibnamefont{Spannowsky}},
  \bibinfo{journal}{Phys. Lett. B} \textbf{\bibinfo{volume}{753}},
  \bibinfo{pages}{482} (\bibinfo{year}{2016}), \eprint{1509.00476}.

\bibitem[{\citenamefont{Gavela et~al.}(2020)\citenamefont{Gavela, No, Sanz, and
  de~Troc\'oniz}}]{Gavela:2019cmq}
\bibinfo{author}{\bibfnamefont{M.~B.} \bibnamefont{Gavela}},
  \bibinfo{author}{\bibfnamefont{J.~M.} \bibnamefont{No}},
  \bibinfo{author}{\bibfnamefont{V.}~\bibnamefont{Sanz}}, \bibnamefont{and}
  \bibinfo{author}{\bibfnamefont{J.~F.} \bibnamefont{de~Troc\'oniz}},
  \bibinfo{journal}{Phys. Rev. Lett.} \textbf{\bibinfo{volume}{124}},
  \bibinfo{pages}{051802} (\bibinfo{year}{2020}), \eprint{1905.12953}.

\bibitem[{\citenamefont{Bauer et~al.}(2017{\natexlab{b}})\citenamefont{Bauer,
  Neubert, and Thamm}}]{Bauer:2017nlg}
\bibinfo{author}{\bibfnamefont{M.}~\bibnamefont{Bauer}},
  \bibinfo{author}{\bibfnamefont{M.}~\bibnamefont{Neubert}}, \bibnamefont{and}
  \bibinfo{author}{\bibfnamefont{A.}~\bibnamefont{Thamm}},
  \bibinfo{journal}{Phys. Rev. Lett.} \textbf{\bibinfo{volume}{119}},
  \bibinfo{pages}{031802} (\bibinfo{year}{2017}{\natexlab{b}}),
  \eprint{1704.08207}.

\bibitem[{\citenamefont{Ellis et~al.}(2013{\natexlab{a}})\citenamefont{Ellis,
  Roy, and Scholtz}}]{Ellis:2012sd}
\bibinfo{author}{\bibfnamefont{S.~D.} \bibnamefont{Ellis}},
  \bibinfo{author}{\bibfnamefont{T.~S.} \bibnamefont{Roy}}, \bibnamefont{and}
  \bibinfo{author}{\bibfnamefont{J.}~\bibnamefont{Scholtz}},
  \bibinfo{journal}{Phys. Rev. Lett.} \textbf{\bibinfo{volume}{110}},
  \bibinfo{pages}{122003} (\bibinfo{year}{2013}{\natexlab{a}}),
  \eprint{1210.1855}.

\bibitem[{\citenamefont{Ellis et~al.}(2013{\natexlab{b}})\citenamefont{Ellis,
  Roy, and Scholtz}}]{Ellis:2012zp}
\bibinfo{author}{\bibfnamefont{S.~D.} \bibnamefont{Ellis}},
  \bibinfo{author}{\bibfnamefont{T.~S.} \bibnamefont{Roy}}, \bibnamefont{and}
  \bibinfo{author}{\bibfnamefont{J.}~\bibnamefont{Scholtz}},
  \bibinfo{journal}{Phys. Rev. D} \textbf{\bibinfo{volume}{87}},
  \bibinfo{pages}{014015} (\bibinfo{year}{2013}{\natexlab{b}}),
  \eprint{1210.3657}.

\bibitem[{\citenamefont{Dobrescu et~al.}(2001)\citenamefont{Dobrescu,
  Landsberg, and Matchev}}]{Dobrescu:2000jt}
\bibinfo{author}{\bibfnamefont{B.~A.} \bibnamefont{Dobrescu}},
  \bibinfo{author}{\bibfnamefont{G.~L.} \bibnamefont{Landsberg}},
  \bibnamefont{and} \bibinfo{author}{\bibfnamefont{K.~T.}
  \bibnamefont{Matchev}}, \bibinfo{journal}{Phys. Rev. D}
  \textbf{\bibinfo{volume}{63}}, \bibinfo{pages}{075003}
  (\bibinfo{year}{2001}), \eprint{hep-ph/0005308}.

\bibitem[{\citenamefont{Chang et~al.}(2007)\citenamefont{Chang, Fox, and
  Weiner}}]{Chang:2006bw}
\bibinfo{author}{\bibfnamefont{S.}~\bibnamefont{Chang}},
  \bibinfo{author}{\bibfnamefont{P.~J.} \bibnamefont{Fox}}, \bibnamefont{and}
  \bibinfo{author}{\bibfnamefont{N.}~\bibnamefont{Weiner}},
  \bibinfo{journal}{Phys. Rev. Lett.} \textbf{\bibinfo{volume}{98}},
  \bibinfo{pages}{111802} (\bibinfo{year}{2007}), \eprint{hep-ph/0608310}.

\bibitem[{\citenamefont{Toro and Yavin}(2012)}]{Toro:2012sv}
\bibinfo{author}{\bibfnamefont{N.}~\bibnamefont{Toro}} \bibnamefont{and}
  \bibinfo{author}{\bibfnamefont{I.}~\bibnamefont{Yavin}},
  \bibinfo{journal}{Phys. Rev. D} \textbf{\bibinfo{volume}{86}},
  \bibinfo{pages}{055005} (\bibinfo{year}{2012}), \eprint{1202.6377}.

\bibitem[{\citenamefont{Draper and McKeen}(2012)}]{Draper:2012xt}
\bibinfo{author}{\bibfnamefont{P.}~\bibnamefont{Draper}} \bibnamefont{and}
  \bibinfo{author}{\bibfnamefont{D.}~\bibnamefont{McKeen}},
  \bibinfo{journal}{Phys. Rev. D} \textbf{\bibinfo{volume}{85}},
  \bibinfo{pages}{115023} (\bibinfo{year}{2012}), \eprint{1204.1061}.

\bibitem[{\citenamefont{Alloul et~al.}(2014)\citenamefont{Alloul, Christensen,
  Degrande, Duhr, and Fuks}}]{Alloul:2013bka}
\bibinfo{author}{\bibfnamefont{A.}~\bibnamefont{Alloul}},
  \bibinfo{author}{\bibfnamefont{N.~D.} \bibnamefont{Christensen}},
  \bibinfo{author}{\bibfnamefont{C.}~\bibnamefont{Degrande}},
  \bibinfo{author}{\bibfnamefont{C.}~\bibnamefont{Duhr}}, \bibnamefont{and}
  \bibinfo{author}{\bibfnamefont{B.}~\bibnamefont{Fuks}},
  \bibinfo{journal}{Comput. Phys. Commun.} \textbf{\bibinfo{volume}{185}},
  \bibinfo{pages}{2250} (\bibinfo{year}{2014}), \eprint{1310.1921}.

\bibitem[{\citenamefont{Alwall et~al.}(2014)\citenamefont{Alwall, Frederix,
  Frixione, Hirschi, Maltoni, Mattelaer, Shao, Stelzer, Torrielli, and
  Zaro}}]{Alwall:2014hca}
\bibinfo{author}{\bibfnamefont{J.}~\bibnamefont{Alwall}},
  \bibinfo{author}{\bibfnamefont{R.}~\bibnamefont{Frederix}},
  \bibinfo{author}{\bibfnamefont{S.}~\bibnamefont{Frixione}},
  \bibinfo{author}{\bibfnamefont{V.}~\bibnamefont{Hirschi}},
  \bibinfo{author}{\bibfnamefont{F.}~\bibnamefont{Maltoni}},
  \bibinfo{author}{\bibfnamefont{O.}~\bibnamefont{Mattelaer}},
  \bibinfo{author}{\bibfnamefont{H.~S.} \bibnamefont{Shao}},
  \bibinfo{author}{\bibfnamefont{T.}~\bibnamefont{Stelzer}},
  \bibinfo{author}{\bibfnamefont{P.}~\bibnamefont{Torrielli}},
  \bibnamefont{and} \bibinfo{author}{\bibfnamefont{M.}~\bibnamefont{Zaro}},
  \bibinfo{journal}{JHEP} \textbf{\bibinfo{volume}{07}}, \bibinfo{pages}{079}
  (\bibinfo{year}{2014}), \eprint{1405.0301}.

\bibitem[{\citenamefont{Sj\"ostrand et~al.}(2015)\citenamefont{Sj\"ostrand,
  Ask, Christiansen, Corke, Desai, Ilten, Mrenna, Prestel, Rasmussen, and
  Skands}}]{Sjostrand:2014zea}
\bibinfo{author}{\bibfnamefont{T.}~\bibnamefont{Sj\"ostrand}},
  \bibinfo{author}{\bibfnamefont{S.}~\bibnamefont{Ask}},
  \bibinfo{author}{\bibfnamefont{J.~R.} \bibnamefont{Christiansen}},
  \bibinfo{author}{\bibfnamefont{R.}~\bibnamefont{Corke}},
  \bibinfo{author}{\bibfnamefont{N.}~\bibnamefont{Desai}},
  \bibinfo{author}{\bibfnamefont{P.}~\bibnamefont{Ilten}},
  \bibinfo{author}{\bibfnamefont{S.}~\bibnamefont{Mrenna}},
  \bibinfo{author}{\bibfnamefont{S.}~\bibnamefont{Prestel}},
  \bibinfo{author}{\bibfnamefont{C.~O.} \bibnamefont{Rasmussen}},
  \bibnamefont{and} \bibinfo{author}{\bibfnamefont{P.~Z.}
  \bibnamefont{Skands}}, \bibinfo{journal}{Comput. Phys. Commun.}
  \textbf{\bibinfo{volume}{191}}, \bibinfo{pages}{159} (\bibinfo{year}{2015}),
  \eprint{1410.3012}.

\bibitem[{\citenamefont{de~Favereau et~al.}(2014)\citenamefont{de~Favereau,
  Delaere, Demin, Giammanco, Lema\^\i{}tre, Mertens, and
  Selvaggi}}]{deFavereau:2013fsa}
\bibinfo{author}{\bibfnamefont{J.}~\bibnamefont{de~Favereau}},
  \bibinfo{author}{\bibfnamefont{C.}~\bibnamefont{Delaere}},
  \bibinfo{author}{\bibfnamefont{P.}~\bibnamefont{Demin}},
  \bibinfo{author}{\bibfnamefont{A.}~\bibnamefont{Giammanco}},
  \bibinfo{author}{\bibfnamefont{V.}~\bibnamefont{Lema\^\i{}tre}},
  \bibinfo{author}{\bibfnamefont{A.}~\bibnamefont{Mertens}}, \bibnamefont{and}
  \bibinfo{author}{\bibfnamefont{M.}~\bibnamefont{Selvaggi}}
  (\bibinfo{collaboration}{DELPHES 3}), \bibinfo{journal}{JHEP}
  \textbf{\bibinfo{volume}{02}}, \bibinfo{pages}{057} (\bibinfo{year}{2014}),
  \eprint{1307.6346}.

\bibitem[{\citenamefont{Cacciari et~al.}(2012)\citenamefont{Cacciari, Salam,
  and Soyez}}]{Cacciari:2011ma}
\bibinfo{author}{\bibfnamefont{M.}~\bibnamefont{Cacciari}},
  \bibinfo{author}{\bibfnamefont{G.~P.} \bibnamefont{Salam}}, \bibnamefont{and}
  \bibinfo{author}{\bibfnamefont{G.}~\bibnamefont{Soyez}},
  \bibinfo{journal}{Eur. Phys. J. C} \textbf{\bibinfo{volume}{72}},
  \bibinfo{pages}{1896} (\bibinfo{year}{2012}), \eprint{1111.6097}.

\bibitem[{CMS(2009)}]{CMS:2009nxa}
 (\bibinfo{year}{2009}).

\bibitem[{\citenamefont{Cacciari et~al.}(2008)\citenamefont{Cacciari, Salam,
  and Soyez}}]{Cacciari:2008gp}
\bibinfo{author}{\bibfnamefont{M.}~\bibnamefont{Cacciari}},
  \bibinfo{author}{\bibfnamefont{G.~P.} \bibnamefont{Salam}}, \bibnamefont{and}
  \bibinfo{author}{\bibfnamefont{G.}~\bibnamefont{Soyez}},
  \bibinfo{journal}{JHEP} \textbf{\bibinfo{volume}{04}}, \bibinfo{pages}{063}
  (\bibinfo{year}{2008}), \eprint{0802.1189}.

\bibitem[{\citenamefont{Catani et~al.}(1993)\citenamefont{Catani, Dokshitzer,
  Seymour, and Webber}}]{Catani:1993hr}
\bibinfo{author}{\bibfnamefont{S.}~\bibnamefont{Catani}},
  \bibinfo{author}{\bibfnamefont{Y.~L.} \bibnamefont{Dokshitzer}},
  \bibinfo{author}{\bibfnamefont{M.~H.} \bibnamefont{Seymour}},
  \bibnamefont{and} \bibinfo{author}{\bibfnamefont{B.~R.}
  \bibnamefont{Webber}}, \bibinfo{journal}{Nucl. Phys. B}
  \textbf{\bibinfo{volume}{406}}, \bibinfo{pages}{187} (\bibinfo{year}{1993}).

\bibitem[{\citenamefont{Ellis and Soper}(1993)}]{Ellis:1993tq}
\bibinfo{author}{\bibfnamefont{S.~D.} \bibnamefont{Ellis}} \bibnamefont{and}
  \bibinfo{author}{\bibfnamefont{D.~E.} \bibnamefont{Soper}},
  \bibinfo{journal}{Phys. Rev. D} \textbf{\bibinfo{volume}{48}},
  \bibinfo{pages}{3160} (\bibinfo{year}{1993}), \eprint{hep-ph/9305266}.

\bibitem[{\citenamefont{Thaler and Van~Tilburg}(2011)}]{Thaler:2010tr}
\bibinfo{author}{\bibfnamefont{J.}~\bibnamefont{Thaler}} \bibnamefont{and}
  \bibinfo{author}{\bibfnamefont{K.}~\bibnamefont{Van~Tilburg}},
  \bibinfo{journal}{JHEP} \textbf{\bibinfo{volume}{03}}, \bibinfo{pages}{015}
  (\bibinfo{year}{2011}), \eprint{1011.2268}.

\bibitem[{\citenamefont{Thaler and Van~Tilburg}(2012)}]{Thaler:2011gf}
\bibinfo{author}{\bibfnamefont{J.}~\bibnamefont{Thaler}} \bibnamefont{and}
  \bibinfo{author}{\bibfnamefont{K.}~\bibnamefont{Van~Tilburg}},
  \bibinfo{journal}{JHEP} \textbf{\bibinfo{volume}{02}}, \bibinfo{pages}{093}
  (\bibinfo{year}{2012}), \eprint{1108.2701}.

\bibitem[{\citenamefont{Butterworth et~al.}(2008)\citenamefont{Butterworth,
  Davison, Rubin, and Salam}}]{Butterworth:2008iy}
\bibinfo{author}{\bibfnamefont{J.~M.} \bibnamefont{Butterworth}},
  \bibinfo{author}{\bibfnamefont{A.~R.} \bibnamefont{Davison}},
  \bibinfo{author}{\bibfnamefont{M.}~\bibnamefont{Rubin}}, \bibnamefont{and}
  \bibinfo{author}{\bibfnamefont{G.~P.} \bibnamefont{Salam}},
  \bibinfo{journal}{Phys. Rev. Lett.} \textbf{\bibinfo{volume}{100}},
  \bibinfo{pages}{242001} (\bibinfo{year}{2008}), \eprint{0802.2470}.

\bibitem[{\citenamefont{Roe et~al.}(2005)\citenamefont{Roe, Yang, Zhu, Liu,
  Stancu, and McGregor}}]{Roe:2004na}
\bibinfo{author}{\bibfnamefont{B.~P.} \bibnamefont{Roe}},
  \bibinfo{author}{\bibfnamefont{H.-J.} \bibnamefont{Yang}},
  \bibinfo{author}{\bibfnamefont{J.}~\bibnamefont{Zhu}},
  \bibinfo{author}{\bibfnamefont{Y.}~\bibnamefont{Liu}},
  \bibinfo{author}{\bibfnamefont{I.}~\bibnamefont{Stancu}}, \bibnamefont{and}
  \bibinfo{author}{\bibfnamefont{G.}~\bibnamefont{McGregor}},
  \bibinfo{journal}{Nucl. Instrum. Meth. A} \textbf{\bibinfo{volume}{543}},
  \bibinfo{pages}{577} (\bibinfo{year}{2005}), \eprint{physics/0408124}.

\bibitem[{\citenamefont{Hocker et~al.}(2007)}]{Hocker:2007ht}
\bibinfo{author}{\bibfnamefont{A.}~\bibnamefont{Hocker}} \bibnamefont{et~al.}
  (\bibinfo{year}{2007}), \eprint{physics/0703039}.

\bibitem[{\citenamefont{Payez et~al.}(2015)\citenamefont{Payez, Evoli, Fischer,
  Giannotti, Mirizzi, and Ringwald}}]{Payez:2014xsa}
\bibinfo{author}{\bibfnamefont{A.}~\bibnamefont{Payez}},
  \bibinfo{author}{\bibfnamefont{C.}~\bibnamefont{Evoli}},
  \bibinfo{author}{\bibfnamefont{T.}~\bibnamefont{Fischer}},
  \bibinfo{author}{\bibfnamefont{M.}~\bibnamefont{Giannotti}},
  \bibinfo{author}{\bibfnamefont{A.}~\bibnamefont{Mirizzi}}, \bibnamefont{and}
  \bibinfo{author}{\bibfnamefont{A.}~\bibnamefont{Ringwald}},
  \bibinfo{journal}{JCAP} \textbf{\bibinfo{volume}{02}}, \bibinfo{pages}{006}
  (\bibinfo{year}{2015}), \eprint{1410.3747}.

\bibitem[{\citenamefont{Jaeckel et~al.}(2018)\citenamefont{Jaeckel, Malta, and
  Redondo}}]{Jaeckel:2017tud}
\bibinfo{author}{\bibfnamefont{J.}~\bibnamefont{Jaeckel}},
  \bibinfo{author}{\bibfnamefont{P.~C.} \bibnamefont{Malta}}, \bibnamefont{and}
  \bibinfo{author}{\bibfnamefont{J.}~\bibnamefont{Redondo}},
  \bibinfo{journal}{Phys. Rev. D} \textbf{\bibinfo{volume}{98}},
  \bibinfo{pages}{055032} (\bibinfo{year}{2018}), \eprint{1702.02964}.

\bibitem[{\citenamefont{D\"obrich et~al.}(2016)\citenamefont{D\"obrich,
  Jaeckel, Kahlhoefer, Ringwald, and Schmidt-Hoberg}}]{Dobrich:2015jyk}
\bibinfo{author}{\bibfnamefont{B.}~\bibnamefont{D\"obrich}},
  \bibinfo{author}{\bibfnamefont{J.}~\bibnamefont{Jaeckel}},
  \bibinfo{author}{\bibfnamefont{F.}~\bibnamefont{Kahlhoefer}},
  \bibinfo{author}{\bibfnamefont{A.}~\bibnamefont{Ringwald}}, \bibnamefont{and}
  \bibinfo{author}{\bibfnamefont{K.}~\bibnamefont{Schmidt-Hoberg}},
  \bibinfo{journal}{JHEP} \textbf{\bibinfo{volume}{02}}, \bibinfo{pages}{018}
  (\bibinfo{year}{2016}), \eprint{1512.03069}.

\bibitem[{\citenamefont{D\"obrich et~al.}(2019)\citenamefont{D\"obrich,
  Jaeckel, and Spadaro}}]{Dobrich:2019dxc}
\bibinfo{author}{\bibfnamefont{B.}~\bibnamefont{D\"obrich}},
  \bibinfo{author}{\bibfnamefont{J.}~\bibnamefont{Jaeckel}}, \bibnamefont{and}
  \bibinfo{author}{\bibfnamefont{T.}~\bibnamefont{Spadaro}},
  \bibinfo{journal}{JHEP} \textbf{\bibinfo{volume}{05}}, \bibinfo{pages}{213}
  (\bibinfo{year}{2019}), \bibinfo{note}{[Erratum: JHEP 10, 046 (2020)]},
  \eprint{1904.02091}.

\bibitem[{\citenamefont{Aad et~al.}(2013)}]{ATLAS:2012fgo}
\bibinfo{author}{\bibfnamefont{G.}~\bibnamefont{Aad}} \bibnamefont{et~al.}
  (\bibinfo{collaboration}{ATLAS}), \bibinfo{journal}{JHEP}
  \textbf{\bibinfo{volume}{01}}, \bibinfo{pages}{086} (\bibinfo{year}{2013}),
  \eprint{1211.1913}.

\bibitem[{\citenamefont{Jaeckel et~al.}(2013)\citenamefont{Jaeckel, Jankowiak,
  and Spannowsky}}]{Jaeckel:2012yz}
\bibinfo{author}{\bibfnamefont{J.}~\bibnamefont{Jaeckel}},
  \bibinfo{author}{\bibfnamefont{M.}~\bibnamefont{Jankowiak}},
  \bibnamefont{and}
  \bibinfo{author}{\bibfnamefont{M.}~\bibnamefont{Spannowsky}},
  \bibinfo{journal}{Phys. Dark Univ.} \textbf{\bibinfo{volume}{2}},
  \bibinfo{pages}{111} (\bibinfo{year}{2013}), \eprint{1212.3620}.

\bibitem[{\citenamefont{Cid~Vidal et~al.}(2019)\citenamefont{Cid~Vidal,
  Mariotti, Redigolo, Sala, and Tobioka}}]{CidVidal:2018blh}
\bibinfo{author}{\bibfnamefont{X.}~\bibnamefont{Cid~Vidal}},
  \bibinfo{author}{\bibfnamefont{A.}~\bibnamefont{Mariotti}},
  \bibinfo{author}{\bibfnamefont{D.}~\bibnamefont{Redigolo}},
  \bibinfo{author}{\bibfnamefont{F.}~\bibnamefont{Sala}}, \bibnamefont{and}
  \bibinfo{author}{\bibfnamefont{K.}~\bibnamefont{Tobioka}},
  \bibinfo{journal}{JHEP} \textbf{\bibinfo{volume}{01}}, \bibinfo{pages}{113}
  (\bibinfo{year}{2019}), \bibinfo{note}{[Erratum: JHEP 06, 141 (2020)]},
  \eprint{1810.09452}.

\bibitem[{\citenamefont{Lees et~al.}(2011)}]{BaBar:2011kau}
\bibinfo{author}{\bibfnamefont{J.~P.} \bibnamefont{Lees}} \bibnamefont{et~al.}
  (\bibinfo{collaboration}{BaBar}), \bibinfo{journal}{Phys. Rev. Lett.}
  \textbf{\bibinfo{volume}{107}}, \bibinfo{pages}{221803}
  (\bibinfo{year}{2011}), \eprint{1108.3549}.

\bibitem[{\citenamefont{Adriani et~al.}(1992)}]{Adriani:1992zm}
\bibinfo{author}{\bibfnamefont{O.}~\bibnamefont{Adriani}} \bibnamefont{et~al.}
  (\bibinfo{collaboration}{L3}), \bibinfo{journal}{Phys. Lett. B}
  \textbf{\bibinfo{volume}{292}}, \bibinfo{pages}{472} (\bibinfo{year}{1992}).

\bibitem[{\citenamefont{Mariotti et~al.}(2018)\citenamefont{Mariotti, Redigolo,
  Sala, and Tobioka}}]{Mariotti:2017vtv}
\bibinfo{author}{\bibfnamefont{A.}~\bibnamefont{Mariotti}},
  \bibinfo{author}{\bibfnamefont{D.}~\bibnamefont{Redigolo}},
  \bibinfo{author}{\bibfnamefont{F.}~\bibnamefont{Sala}}, \bibnamefont{and}
  \bibinfo{author}{\bibfnamefont{K.}~\bibnamefont{Tobioka}},
  \bibinfo{journal}{Phys. Lett. B} \textbf{\bibinfo{volume}{783}},
  \bibinfo{pages}{13} (\bibinfo{year}{2018}), \eprint{1710.01743}.

\end{thebibliography}

\end{document}